\documentclass[utf8]{frontiersSCNS} % for Science, Engineering and Humanities and Social Sciences articles

\setcitestyle{square} % for Physics and Applied Mathematics and Statistics articles
\usepackage{url,hyperref,lineno,microtype,subcaption}
\usepackage[onehalfspacing]{setspace}

\usepackage{titlesec, blindtext, color}
	\definecolor{light-blue}{rgb}{0.8,0.85,1}
	\definecolor{airforceblue}{rgb}{0.36, 0.54, 0.66}
	\definecolor{azure}{rgb}{0.0, 0.5, 1.0}
	\definecolor{bleudefrance}{rgb}{0.19, 0.55, 0.91}
	\definecolor{blue(munsell)}{rgb}{0.0, 0.5, 0.69}
	\definecolor{darkmidnightblue}{rgb}{0.0, 0.2, 0.4}
	\definecolor{steelblue}{rgb}{0.27, 0.51, 0.71}
	\definecolor{tealblue}{rgb}{0.21, 0.46, 0.53}
	\definecolor{yaleblue}{rgb}{0.06, 0.3, 0.57}
	\definecolor{applered}{rgb}{0.89, 0.02, 0.17}

\newcommand{\radyn}{\texttt{RADYN}}
\newcommand{\fp}{\texttt{FP}}

\newcommand{\radynarcade}{\texttt{RADYN\_Arcade}}
\newcommand{\msradyn}{\texttt{MS\_RADYN}}
\newcommand{\deft}{\texttt{DEFT}}
\newcommand{\preft}{\texttt{PREFT}}
\newcommand{\hydrad}{\texttt{HYDRAD}}
\newcommand{\flarix}{\texttt{FLARIX}}
\newcommand{\hydrotwogen}{\texttt{HYDRO2GEN}}
\newcommand{\rh}{\texttt{RH}}
\newcommand{\rhpar}{\texttt{RH15D}}
\newcommand{\mali}{\texttt{MALI}}

\newcommand\ion[2]{#1$\;${%
\ifx\@currsize\normalsize\small \else
\ifx\@currsize\small\footnotesize \else
\ifx\@currsize\footnotesize\scriptsize \else
\ifx\@currsize\scriptsize\tiny \else
\ifx\@currsize\large\normalsize \else
\ifx\@currsize\Large\large
\fi\fi\fi\fi\fi\fi
\rmfamily\textsc{#2}}\relax}% 

\newcommand{\gskfont}{
  \bfseries
  \color{applered}
}
\DeclareTextFontCommand{\gsk}{\gskfont}

\def\keyFont{\fontsize{8}{11}\helveticabold }
\def\firstAuthorLast{Kerr, G.S.} 
\def\Authors{Graham S. Kerr\,$^{1,2,*}$}
\def\Address{$^{1}$Department of Physics, The Catholic Univeristy of America, 620 Michigan Avenue, Northeast, Washington D.C. 20064, USA \\ 
$^{2}$NASA Goddard Space Flight Center, Heliophysics Sciences Division, Code 671, 8800 Greenbelt Road, Greenbelt MD 20771, USA }

\def\corrAuthor{Graham S. Kerr}

\def\corrEmail{graham.s.kerr@nasa.gov, kerrg@cua.edu}

\begin{document}
\onecolumn
\firstpage{1}

\title[Solar Flare Loop Models \& IRIS, Paper 1]{Interrogating Solar Flare Loop Models with IRIS Observations 1: Overview of the Models, and Mass flows} 

\author[\firstAuthorLast ]{\Authors} %This field will be automatically populated
\address{} %This field will be automatically populated
\correspondance{} %This field will be automatically populated

\extraAuth{}% If there are more than 1 corresponding author, comment this line and uncomment the next one.
%\extraAuth{}

\maketitle

%%%%%%%%%%%%%%%%%%%%%%%%%%%%%%%%%%%%%%%%%%%%%
%% ABSTRACT
%%%%%%%%%%%%%%%%%%%%%%%%%%%%%%%%%%%%%%%%%%%%%

\begin{abstract}

\section{}
Solar flares are transient yet dramatic events in the atmosphere of the Sun, during which a vast amount of magnetic energy is liberated. This energy is subsequently transported through the solar atmosphere or into the heliosphere, and together with coronal mass ejections flares comprise a fundamental component of space weather. Thus, understanding the physical processes at play in flares is vital. That understanding often requires the use of forward modelling in order to predict the hydrodynamic and radiative response of the solar atmosphere. Those predictions must then be critiqued by observations to show us where our models are missing ingredients. While flares are of course 3D phenomenon, simulating the flaring atmosphere including an accurate chromosphere with the required spatial scales in 3D is largely beyond current computational capabilities, and certainly performing parameter studies of energy transport mechanisms is not yet tractable in 3D. Therefore, field-aligned 1D loop models that can resolve the relevant scales have a crucial role to play in advancing our knowledge of flares. In recent years, driven in part by the spectacular observations from the Interface Region Imaging Spectrograph (IRIS), flare loop models have revealed many interesting features of flares. For this review I highlight some important results that illustrate the utility of attacking the problem of solar flares with a combination of high quality observations, and state-of-the-art flare loop models, demonstrating: (1) how models help to interpret flare observations from IRIS, (2) how those observations show us where we are missing physics from our models, and (3) how the ever increasing quality of solar observations drives model improvements. Here in Paper 1 of this two part review I provide an overview of modern flare loop models, and of electron-beam driven mass flows during solar flares. 

\tiny
 \keyFont{ \section{Keywords:} solar flares, solar atmosphere, solar chromosphere, UV radiation, Numerical methods, Radiation Transfer} %All article types: you may provide up to 8 keywords; at least 5 are mandatory.
\end{abstract}

%%%%%%%%%%%%%%%%%%%%%%%%%%%%%%%%%%%%%%%%%%%%%
%% INTRODUCTION
%%%%%%%%%%%%%%%%%%%%%%%%%%%%%%%%%%%%%%%%%%%%%

\section{Introduction}
Understanding the physical mechanisms responsible for, and at play during, solar flares still remains one of the most important open issues in astrophysics. These energetic events release a tremendous amount of magnetic energy, which can be $>10^{32}$~ergs, resulting in efficient particle acceleration and are often associated with the ejection of coronal material \citep[e.g.][]{2012ApJ...759...71E}. Flares and coronal mass ejections (CMEs), together solar eruptive events (SEEs), both strongly influence space weather making understanding flare processes a practical necessity as well as an interesting scientific problem. It is generally believed that the stressed magnetic field reconfigures via magnetic reconnection, liberating energy \citep[e.g.][]{2002A&ARv..10..313P, 2011LRSP....8....6S,2013A&A...555A..77J}. The primary energy release site is the solar corona, though it is possible that energy release also takes place elsewhere. Following reconnection, this energy manifests in various physical forms, such as the acceleration of vast numbers of particles (up to $10^{36}$ electrons per second!), direct heating of the corona, mass flows, coronal mass ejections (CMEs), and possibly magnetohydrodynamic (MHD) waves. 

Ultimately, a significant fraction of this energy is transported to the lower solar atmosphere, that is the transition region, chromosphere, and potentially even deeper to the temperature minimum region and photosphere. Intense heating and ionisation occurs, producing the broadband enhancement to the Sun's radiative output that characterises the flare \citep[e.g.][]{2008LRSP....5....1B,2011SSRv..159...19F}. An expansion of chromospheric layers occurs, driving mass flows both upwards into the corona (chromospheric evaporation) filling it with chromospheric material, and downwards to deeper layers (chromospheric condensation). The mass-loaded and heated coronal loops subsequently emit strongly, forming bright flare loops that often appear as part of a large scale arcade structure due to propagation of magnetic reconnection. The strength of a flare is defined by the flux of soft X-rays (primarily emitted from flare loops), as observed by the one-minute data from the X-ray Sensor B (1-8\AA) on board NOAA's Geostationary Operational Environmental Satellite (GOES/XRSB) satellites. On a logarithmic scale flares are classified as [A, B, C, M, X], from weakest to strongest, with sub-divisions of, e.g. M1-10. Sub-A class \cite{Glesener_2017,2021MNRAS.507.3936C} flares have been observed, as have flares $>$X10 \citep[e.g.][]{2012ApJ...759...71E}.

In the standard model of solar flares the dominant means by which energy is transported from the coronal release site to the chromosphere and transition region is thought to be by via directed beams of non-thermal particles, accelerated out of the thermal background. It is common to mainly consider non-thermal electrons in flare models given the volume of evidence of their presence in the lower atmosphere, but comparable energies (roughly within an order of magnitude) in flare accelerated protons have been observed \citep[][]{2012ApJ...759...71E}. Due in large part to the lack of physical constraints on the distribution of those protons they are typically omitted in flare models, and we concentrate primarily on electrons, however there is evidence in three flares of energetic protons near flare ribbons \citep[][]{2003ApJ...595L..77H,2006ApJ...644L..93H}. Once in the denser chromosphere these energetic particles undergo Coulomb collisions, thermalising and heating the plasma, accompanied by the production of hard X-rays via bremsstrahlung \citep[e.g.][]{1971SoPh...18..489B,1978ApJ...224..241E}. A substantial body of evidence supports the important role that non-thermal electrons play in transporting flare energy, with a great many observations of hard X-rays, for example from the Reuven Ramaty High Energy Solar Spectroscopic Imager \citep[RHESSI;][]{2002SoPh..210....3L}, that are co-spatial and co-temporal with other flare radiation \citep[e.g.][]{2011SSRv..159...19F,2011ApJ...739...96K}. From inversions of the X-ray energy spectrum it is possible to infer the spectral properties of the non-thermal electrons that bombard the chromosphere \citep[see reviews by][]{2011SSRv..159..107H,2011SSRv..159..301K}, which can subsequently be used to drive flare models of the type discussed in this review. It should be noted that there are caveats to this process, which can lead to uncertainties in the inferred non-thermal electron spectral properties. Uncertainties can be due to model assumptions \citep[e.g. ignoring warm target or return current effects,][]{2015ApJ...809...35K,2019ApJ...880..136J,2017ApJ...851...78A,2020ApJ...902...16A}, or due to the particular difficulty in obtaining reliable estimates of the low-energy cutoff, $E_{c}$ (where the spectrum transitions from thermal to non-thermal). In most fitting procedures this low-energy cutoff is taken to be the largest value consistent with the data (e.g. $\chi_{red} \sim 1$) but, since the thermal emission masks the non-thermal emission at these small energies, $E_{c}$ could in fact be much smaller, hence the derived estimate of the power carried by non-thermal electrons is essentially a lower-limit \citep[see, for example, discussions in][]{2011SSRv..159..107H,2011SSRv..159..301K,2012ApJ...759...71E,2015ApJ...809...35K,2016A&A...588A.116W,2020A&A...644A.172W,2021ApJ...917...74A}. The energy spectrum has an assumed power-law form, the parameters of which are generally the spectral index $\delta$ describing the slope of the power-law, and the total energy flux $F$, above some low-energy cutoff $E_{c}$ and below some break energy. See the following reviews for in-depth discussions of the so-called `electron-beam' model: \cite{2011SSRv..159..107H,2011SSRv..159..301K,2011SSRv..159..357Z}. Radio and microwaves can also give powerful diagnostics of non-thermal particles, for example recent studies using the Expanded Owens Valley Solar Array \citep[EOVSA;][]{2018ApJ...863...83G,2020Sci...367..278F,2022Natur.606..674F,2020ApJ...895L..50C,2020NatAs...4.1140C}. Other mechanisms of energy transport in flares include non-thermal protons or heavier ions, thermal conduction following direct heating of the corona, and Alfv\'enic waves, though it is not yet known under which circumstances each mechanism plays a significant role compared to the typically modelled non-thermal electrons. These are are discussed in Paper 2 of this review \citep[][]{kerr2022_irisflarereview_p2}. 

The flare impulsive phase describes the rapid release and deposition of energy that generally lasts a few minutes up to tens of minutes, and which is usually associated with the detection of hard X-rays. The flare gradual phase is the period during which flare emissions, as the moniker implies, gradually decrease and the flare plasma cools. This takes place over tens of minutes, or even hours in some long duration events. Due to the fact that flare models predict much shorter cooling timescales than are observed, and that there is evidence of late-phase evaporation \citep[see e.g.][]{1999ApJ...521L..75C,2001ApJ...552..849C}, there have been strong suggestions that there is some post-impulsive phase (i.e. in the absence of hard X-ray footpoint emission) energy release of unknown form, perhaps even rivalling that of the impulsive phase \citep[see discussions in][]{2016ApJ...820...14Q,2017ApJ...835....6K,2018ApJ...856...27Z,2018ApJ...865...67E,2022ApJ...931...60A}. There is evidence of yet further energy release up to several hours after the traditional gradual phase in some events, known as the `EUV late-phase' owing to their identification in EUV data \citep[][]{2011ApJ...739...59W,2014SoPh..289.3391W}, though they have since been studied in X-rays also \citep[][]{2017ApJ...835....6K}. However, since this review concerns IRIS observations and flare loop models, not the global structure that might be responsible for the EUV late-phase \citep{2014SoPh..289.3391W}, I focus mostly on the impulsive phase footpoints. 

Flare emission can appear variously as compact kernels or footpoints (e.g. white light continua enhancements, hard X-rays, microwaves), extended ribbon-like structures (infrared, optical, UV, extreme-UV), or along the legs of coronal loops (EUV, soft X-rays). Looptop sources of hard X-rays, radio and microwaves are also observed, indicating populations of both very hot (up to tens of MK) thermal plasma, and non-thermal particles. The bulk of this review will focus on modelling of flare footpoints and ribbons, and what we can learn from that emission about flare energy transport and deposition mechanisms. I direct readers to the following detailed reviews of flare observations for a more general overview: \cite{2008LRSP....5....1B,2011SSRv..159...19F,2011SSRv..159..107H,2015SoPh..290.3399M}.

The chromosphere and transition region, as well as being the locations where the bulk of flare energy is deposited, are where the bulk of the flare enhanced radiative output originates, and are thus excellent sources of diagnostic potential. However, the chromosphere and transition region are exceptionally complex environments, particularly so during dynamic events like flares. They are regions with strong gradients in temperature, density and velocity, that are partially ionised with a transition to being fully ionised over what can be vanishingly short distances. Further, the radiation field plays a significant role in plasma heating and cooling, and in spectral line formation, such that non-local thermodynamic (NLTE) effects are present. Observations from the Interface Region Imaging Spectrograph \citep[IRIS;][]{2014SoPh..289.2733D} now provide an unprecedented view of the flaring chromosphere and transition region, yielding crucial new insights.

IRIS is a NASA Small Explorer mission that since its launch in 2013 has observed many hundreds of flares, including dozens of M and X class events, in the far-, and near-UV (FUV \& NUV) offering a new window on the solar chromosphere, and transition region as well as hot flaring plasma via the \ion{Fe}{xxi} 1354.1~\AA\ line. It is a slit scanning spectrograph, that offers high spatial resolution ($0.33-0.4$~arcseconds) spectra, at high cadences of a few to tens of seconds, but to down $1$~s in some events, from a slit $175$~arcseconds in length. IRIS can operate in either a sit-and-stare mode, or can raster over a field of view (FOV), so that the full FOV possible is $130\times175$~arcseconds. Observations probe many layers of the chromosphere and transition region in three passbands: $[1332-1358]$~\AA, $[1389-1407]$~\AA, and $[2783-2834]$~\AA, though it is rare to have full readout, with subsets of lines selected instead. The strongest lines are: \ion{Mg}{ii} h 2803~\AA\ \& k 2796~\AA\  (chromosphere), \ion{C}{ii} 1334~\AA\ \& 1335~\AA\ and \ion{Si}{iv} 1394~\AA\ \& 1403~\AA\ (transition region), and \ion{Fe}{xxi} 1354.1~\AA\ ($\sim11$~MK plasma), with numerous other weaker lines such as \ion{O}{i} 1355.6~\AA, lines of \ion{O}{iv}, lines of \ion{Fe}{ii}, as well as lines of singly ionised species and even molecular H2 transitions. These lines are observable with resolutions of $\sim53$~m\AA\ in the NUV and $\sim26$~m\AA\ in the FUV. Alongside the spectrograph (SG) there is a slit-jaw imager (SJI) with four passbands available up to $175\times175$~arcseconds FOV at $2796\pm4$~\AA\ (\ion{Mg}{ii} k), $2832\pm4$~\AA\ (\ion{Mg}{ii} wing plus quasi-continuum), $1330\pm55$~\AA\ (\ion{C}{ii}), and $1400\pm55$~\AA\ (\ion{Si}{iv}). IRIS has been an extraordinary mission that has brought about a renewed interest in chromospheric studies of both flare and quiescent phenomenon. A recent review provides an overview of the mission's successes: \cite{2021SoPh..296...84D}.

Major observational advancements can only be fully exploited if there is a parallel development and improvement of the theoretical models which are used to interpret those observations. This is particularly true when modelling the optically thick emission from the lower atmosphere, requiring advanced radiative transfer calculations, as well as the treatment of non-equilibrium conditions in the tenuous optically thin corona during dynamic events like flares. State-of-the-art modelling of the flare chromosphere and transition region is required to fully appreciate the information that the observations convey. In this review I discuss the interplay between recent chromospheric and transition region observations from IRIS, and flare loop modelling (with some digressions to coronal emission, mostly in the context of chromospheric evaporation). I demonstrate: (1) how modelling has helped interpret the IRIS observations; (2) how IRIS observations have been used to interrogate and validate model predictions; and (3) how, when models fail to stand up to the stubborn reality of those observations, IRIS has led to model improvements. This review is in two parts. In this Paper 1 I discuss the codes themselves and flare-induced mass flows, and I discuss plasma properties, energy transport mechanisms, and future directions in Paper 2 \citep{kerr2022_irisflarereview_p2}.

Field-aligned, 1D, (radiation-) hydrodynamic models are now routinely used to study the atmospheric plasma response to the heating in an individual flare loop. The advantage of such models is that they allow us to simulate the plasma dynamics at very small spatial scales. It is often required to resolve down to sub-metre scales due to sharp gradients and shocks that form following the injection of flare energy. It is also very important to adequately resolve the transition region, even in quiescent scenarios \citep[e.g. see discussions in][]{2017A&A...597A..81J,2017A&A...605A...8J}, the exceptionally narrow interface between the flaring chromosphere and corona. Achieving the required temporal and spatial resolution for a flare simulation in a 2D or 3D model that includes an accurate NLTE chromosphere with appropriate radiative heating and cooling would be very computationally demanding. The 1D assumption is justified by the fact that in the low plasma $\beta$ regime of the solar atmosphere, mass and energy transport across the magnetic field is highly inhibited, and it is therefore appropriate to treat each flare strand as an isolated plasma loop. Further, since they are much more computationally tractable, field-aligned models allow us to perform large parameter studies of flares driven by different energy transport mechanisms on reasonable timescales, which include the appropriate physical processes. It is essential that we understand the complex physics involved in a field-aligned model before progressing to 3D. There have now been 3D RMHD codes that have modelled the build up and release of flare energy, and subsequent atmospheric heating \citep[e.g.][]{2019NatAs...3..160C}. While impressive feats that give invaluable insight to flare energy release, those models do not yet include energetic particles, nor do they model chromospheres in as much as detail as the loop models discussed in this review. Also, large parameter studies of energy transport processes are currently precluded by the computational demands of 3D RMHD simulations.

The models discussed in detail in this review are all modern numerical codes that are now well-established but which have a rich heritage built upon efforts dating from the 1980s. I do not intend to provide an exhaustive list of historical field-aligned models, but direct the reader to consult the following literature, and references therein: \cite{1980ApJ...239.1036C}, \cite{1982PhDT.........3R}, \cite{1983ApJ...265.1090C}, \cite{1983ApJ...272..739R}, \cite{1983ApJ...265..483M}, \cite{1984ApJ...282..296C}, \cite{1985ApJ...289..414F,1985ApJ...289..425F,1985ApJ...289..434F,1989ApJ...346.1019F}, \cite{1986SoPh..103...47M}, \cite{1984MmSAI..55..811K}.

%%%%%%%%%%%%%%%%%%%%%%%%%%%%%%%%%%%%%%%%%%%%%
%% LOOP MODEL SUMMARY
%%%%%%%%%%%%%%%%%%%%%%%%%%%%%%%%%%%%%%%%%%%%%

\section{Overview of Modern Flare Loop Models}\label{sec:models}
\begin{figure}[h]
\begin{center}
\includegraphics[width=\textwidth]{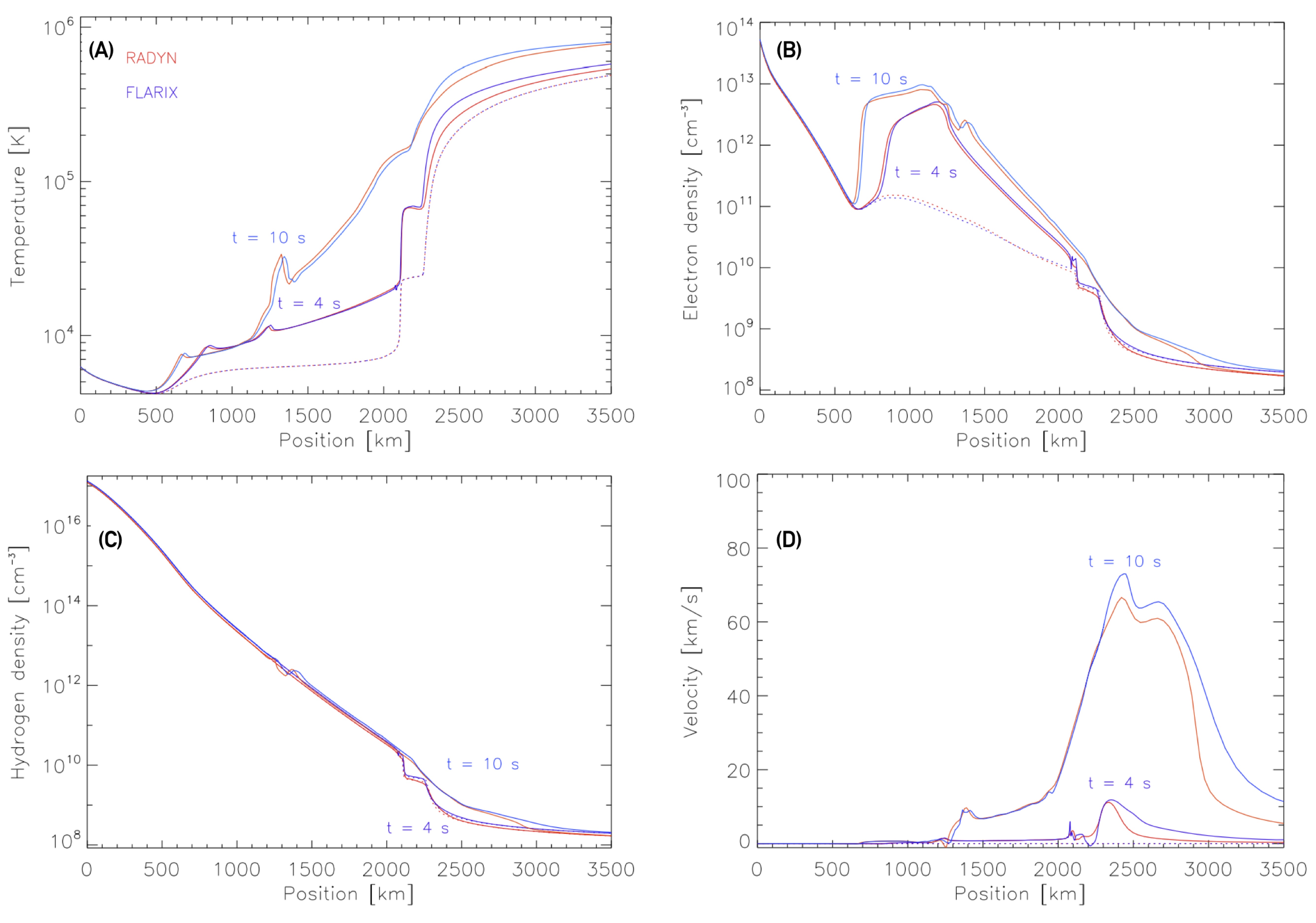}% This is a *.eps file
\end{center}
\caption{A comparison between \radyn\ and \flarix, in which each code was purposefully stripped down to include as similar physics as possible. Despite very different numerical schemes, both codes produced strikingly similar results following injection of an electron beam. In each panel red is \radyn, blue is \flarix, and the dotted lines are  $t=0$~s. Two snapshots during the flare are shown, at $t=4$~s and $t=10$~s. Panel \textbf{(A)} is temperature, \textbf{(B)} is electron density, \textbf{(C)} is hydrogen density, and \textbf{(D)} is velocity (upflows are positive). Figure adapted from \cite{kaspraova2019}.} \label{fig:radynflarixcomp}
\end{figure}

Here I introduce some flare loop models that have been used alongside IRIS data, namely: \radyn, \hydrad, \flarix, \& \preft. There are other flare loop models either currently in use, or that laid foundations, but which have not been used in conjunction with IRIS observations so are outwith the scope of this particular review. In particular I would like to draw the reader's attention to \hydrotwogen, which has been used to study hydrogen line and continua emission in flares \citep[e.g.][]{2017NatCo...815905D,2018A&A...610A..68D,2019A&A...623A..20D}.  A natural question at the outset is `how well do these codes compare?' Each code has very different numerical schemes and approaches, but efforts to compare their predictions have shown that the flaring hydrodynamic response between \radyn\ and \flarix\ is strikingly similar \citep[][]{kaspraova2019}! In that test the codes were stripped down to include as similar physics as possible, so that any major differences present were mostly due to the numerical approach. Figure~\ref{fig:radynflarixcomp} shows the hydrodynamic variables at two snapshots from each code, illustrating their similarities, and that differences were relatively minor. Efforts to compare in detail the radiative predictions, and also to compare the predictions from \hydrad\ to those from \radyn\ and \flarix\ are actively underway as part of an International Space Science Institute team, with promising results thus far.  

\subsection{RADYN}
\radyn\ \citep[][]{1992ApJ...397L..59C,2002ApJ...572..626C,1997ApJ...481..500C,1999ApJ...521..906A,2005ApJ...630..573A,2015ApJ...809..104A} is a radiation hydrodynamic (RHD) code (written in Fortran) that solves the coupled nonlinear equations of hydrodynamics, charge conservation, time-dependent (non-equilibrium) atomic level populations, and radiation transfer on a 1D field-aligned adaptive grid \citep[][]{1987JCoPh..69..175D}. This adaptive scheme allows \radyn\ to resolve the strong shocks and gradients that usually form in flare simulations, and typically has 191 or 300 grid points (though this changeable). A semi-circular loop geometry is assumed, with one half of a symmetric flux tube modelled, and a reflecting boundary condition at the loop apex designed to mimic incoming disturbances from the other half of the loop. This loop spans the sub-photosphere, through the chromosphere, transition region, and corona. Equations are solved in the linearised form using a fully implicit scheme \citep{1998PhDT.........1A}.

Species important for chromospheric energy balance are computed in detail, solving the NLTE radiation transfer and atomic level populations with the methods of \cite{1981ApJ...249..720S} and \cite{1985JCoPh..59...56S}. Those species are a six-level-with-continuum \ion{H}{i} atom, a six-level-with-continuum \ion{Ca}{ii} ion, and a nine-level-with-continuum helium atom/ion (with transitions of \ion{He}{i} and \ion{He}{ii}). In some models a \ion{Mg}{ii} ion is also included. See \cite{2015ApJ...809..104A} for a list of the typical bound-bound and bound-free transitions. Bound-bound transitions are computed assuming complete frequency redistribution (CRD)\footnote{In brief, CRD assumes that the wavelength of a scattered/emitted photon is uncorrelated to the wavelength at which it was absorbed, due to collisions (e.g. photons absorbed in the line wings may be redistributed and emitted at a wavelength in the line core). However, in relatively low-density environments such as the chromosphere there may be an insufficient number of elastic collisions such that the scattered photon has a wavelength that is correlated to that of the absorbed photon. Photons absorbed in the line wings are re-emitted in the wings, where it easier to escape. This is the partial frequency redistribution (PRD) scenario. CRD has a frequency independent source function, whereas PRD has a frequency dependent source function and the absorption profile does not equal the emission profile. See discussions in \cite{1982JQSRT..27..593H}, and \cite{2001ApJ...557..389U,2002ApJ...565.1312U}.}. To avoid overestimating radiative losses from the line wings, the Lyman lines mimic partial frequency redistribution (PRD) by either truncating at 10 Doppler widths, or modelling the line as a pure Doppler profile, depending on which version of the code is being used.  Other species are included as a source of background continuum opacity via the Uppsala opacity package \citep{1973UppAn...5f...1G}. Optically thin losses are included by summing transitions from the CHIANTI atomic database \cite[][]{1997A&AS..125..149D,2015A&A...582A..56D}, excluding those transitions solved in detail. Downward directed incident radiation is included in the solution of the radiation transfer equation, so that photoionisations from X-ray, EUV and UV radiation are considered. This is achieved by calculating the sum of emissivities from transitions in CHIANTI, using the local temperature and density within each grid cell. Thermal conduction is a modified form of Spitzer conductivity, that saturates at the free-streaming limit, though, \cite{2022ApJ...931...60A} added the option to suppress thermal conduction using the method of \cite{2018ApJ...865...67E}, which accounts for turbulence or non-local effects.

Flares are typically simulated by injecting a beam of non-thermal electrons at the apex of the loop, which are then thermalised, heating the plasma. Pre-2015 this was achieved using the analytic expressions of \cite{1978ApJ...224..241E} and \cite{1994ApJ...426..387H}, but post \cite{2015ApJ...809..104A} this is achieved using Fokker-Planck kinetic theory, following \cite{1990ApJ...359..524M}, that better captures scattering terms, and which is applicable no-matter the target temperature (that is, there is no need to make a cold or warm target assumption, both are modelled using the actual target temperature). More recently still, \cite{2020ApJ...902...16A} developed the standalone open-source \fp \footnote{\url{https://github.com/solarFP/FP}} code to more accurately solve the non-thermal particle transport and energy dissipation, including the ability to include beam-neutralising return current effects, and to model the transport of non-thermal protons. \fp\ has now been merged with \radyn. In all cases, non-thermal collisional ionisations and excitations of hydrogen by the particle beams are included, using the the \cite{1993A&A...274..917F} approach, and post-\cite{2015ApJ...809..104A}, non-thermal ionisation of helium is included via data from \cite{1985A&AS...60..425A}. Other options allow \radyn\ to model flare energy transport by mono-chromatic Alfv\'enic waves \citep{2016ApJ...827..101K} or by \textsl{ad-hoc} time-dependent heating.  

\texttt{RADYN} also allows us to calculate \textsl{a posteriori} (i.e. with no feedback on the plasma equations of mass, momentum, and energy) the time-dependent (non-equilibrium) populations and radiation transport of a desired ion via the minority species version of that code, \msradyn\ \citep[][]{2003ApJ...597.1158J,2019ApJ...885..119K,2019ApJ...871...23K}. In this manner the hydrodynamic variables at each internal \radyn\ timestep, written separately from the main output file (that may have too low a cadence), can be used to calculate any additional species including non-equilibrium effects.

\subsection{HYDRAD}
The HYDrodynamics and RADiation (\hydrad) code was originally developed to model the field-aligned plasma physics of solar coronal loops subject to impulsive thermal heating \cite{2003A&A...401..699B,2003A&A...407.1127B}.  Particularly careful attention is paid to the time-dependent (non-equilibrium) evolution of any desired ion species and their radiative coupling to the plasma, and to dynamically capturing the small spatial scales that arise in the solar transition region. 

\hydrad\ solves the conservative form (mass, momentum, and energy density) of the hydrodynamic equations for a two-fluid plasma, on a grid that employs adaptive mesh refinement of arbitrary order. The loops can have any geometry, length, inclination, or cross-section, and span footpoint-to-footpoint (for flare runs, to date), or as an open field line configuration \citep[e.g.][]{2022ApJ...933...72S}, with a corona, transition region, and stratified chromosphere. Prior to \cite{2019ApJ...871...18R} the chromospheric ionisation fraction was calculated with LTE assumptions, but \cite{2019ApJ...871...18R} implemented an approach that aims to capture NTLE hydrogen effects by approximating the radiation field without solving the full radiation transfer problem. Radiative losses in the chromosphere make use of the lookup tables of \cite{2012A&A...539A..39C}, which account for losses from hydrogen, calcium and magnesium. Coronal radiative losses are calculated by summing the emissivity of all transitions within the CHIANTI database, as a function of the ion population fraction, where the ionisation state can be given in equilibrium or calculated out-of-equilibrium, and the emission measure in each grid cell.

Flares are simulated by injecting a power-law distribution of non-thermal electrons at the loop apex, following the analytic treatment of \cite{1978ApJ...224..241E} and \cite{1994ApJ...426..387H}, with a sharp low-energy cutoff. This was implemented in \cite{2013ApJ...778...76R}, and non-thermal collisional ionisation and excitations of hydrogen were added in \cite{2019ApJ...871...18R} using the \cite{1993A&A...274..917F} expressions for those rates. Flares driven by mono-chromatic (i.e. a single frequency) Alfv\'enic waves have also be modelled \citep{2016ApJ...818L..20R,2018ApJ...853..101R}.

Over the years \hydrad\ has evolved into a flexible and powerful code capable of modelling a broad variety of phenomena including: multi-species plasma confined to full-length, magnetic flux tubes of arbitrary geometrical and cross-section variation in the field-aligned direction \citep{2016ApJ...821...63B}; solar flares driven by non-thermal electrons and mono-chromatic Alfv\'enic waves, and the non-equilibrium response of the chromosphere; coronal rain formed by condensations in thermal non-equilibrium where the adaptive grid is required to fully resolve and track multiple steep transition regions \citep{2019A&A...625A.149J}; and ultracold, strongly coupled laboratory plasmas,  composed of weakly-ionised strontium \citep{2013PhPl...20d3516M,2015PhPl...22d3514M}.

\hydrad\ is written in C++ and is designed to be modular in its structure, such that new capabilities (e.g. physical processes) can be added in a relatively straightforward way and handled robustly by the numerical scheme. It is also intended to be fairly undemanding of computational resources, though its needs do depend strongly on the particular nature of each model run (e.g. physics requirements, spatial resolution). The recently implemented NLTE solver for a 6-level hydrogen atom in the optically-thick chromosphere necessitated parallelisation of part of the code (the OpenMP standard is employed) to recover acceptable runtimes. A significant performance gain may also be obtained when solving for the time-dependent ionisation state of a large number of elements coupled to the electron energy equation via the radiative loss term. Otherwise, if this functionality is not required, then \hydrad\ is generally most efficiently executed in single-processor mode with multiple instances running in an ``embarrassingly-parallel'' exploration of a parameter space, for example.

The code has been extensively deployed, tested, and used for a large number of scientific investigations on Windows PC, Mac, and Linux platforms, and found to be stable and robust. \hydrad\ can be freely downloaded from its GitHub repository\footnote{\url{https://github.com/rice-solar-physics/hydrad}}.

\subsection{\textsl{Flarix}}
\flarix\ is a hybrid radiation hydrodynamic code (written in Fortran), comprised of three parts \citep{2016IAUS..320..233H}. Each component can be run as standalone codes, but are fully integrated within \flarix. They are (1) a test-particle code that models the transport and thermalisation of non-thermal particles \citep[][]{2010ITPS...38.2249V,2014A&A...563A..51V}, (2) a 1D field-aligned hydrodynamic code \citep[e.g.][]{2009A&A...499..923K}, and (3) a time-dependent (non-equilibrium) NLTE radiative transfer code \citep[\mali; e.g.][]{1995A&A...299..563H,2003ASPC..288..544K}. \flarix\ solves the hydrodynamics and NLTE radiation transport equations separately, but with feedback between the two codes so that, like \radyn, radiative heating and cooling from chromospheric lines and continua are considered, as is an accurate time-dependent NLTE hydrogen ionisation fraction. 

\flarix\ solves the single fluid hydrodynamic equations along one leg of a symmetric magnetic loop, that is assumed to be semi-circular. When solving those equations the time-dependent hydrogen ionisation fraction is obtained from the NLTE radiation transport code \mali, with the coronal segment assumed to be fully ionised (the ionisation fraction explicitly set to 1). The conductive heat flux is the Spitzer classical formula, and mechanical heating is applied to assure stability of the pre-flare atmosphere, which is typically a VAL-C \citep{1981ApJS...45..635V} type stratification. This atmosphere spans the sub-photosphere through corona, with a fixed grid of $\sim2000$ cells that is optimised to resolve gradients and shocks in the flare chromosphere and transition region. Radiative losses in the chromosphere are computed at each timestep using \mali\ with the instantaneous values of temperature, electron density, hydrogen density, and non-thermal electron heating rate (to account for non-thermal collisional rates) from the hydrodynamics piece. The full plane-parallel radiation transfer problem is solved for appropriate bound-bound and bound-free transitions of \ion{H}{i}, \ion{Ca}{ii}, and \ion{Mg}{ii} (with the addition of helium in progress), assuming CRD (with Lyman lines truncated at 10 Doppler widths), and ensuring charge and particle conservation. In the coronal part of the loop radiative losses are assumed optically thin, employing the loss function of \cite{1978ApJ...220..643R}. 

Flares are simulated by injecting a distribution of non-thermal electrons or protons, which are propagated and thermalised by Coulomb collisions (subsequently heating the plasma) using test particle and Monte Carlo methods, following the approach of \cite{1982ApJ...259..341B} and \cite{1992A&A...264..679K}. This includes the relevant scattering terms, and pitch angle effects, and is equivalent to solving directly the Fokker-Planck equations \citep[][]{1991A&A...251..693M}, but also provides a flexible means to investigate many aspects of non-thermal electron or proton interactions, such as magnetic mirroring and return current effects \citep[e.g.][]{2014A&A...563A..51V}. Alternatively, the analytic expressions of \cite{1978ApJ...224..241E} and \cite{1994ApJ...426..387H} can also be used.  Non-thermal rates follow the \cite{1993A&A...274..917F} approach. For full details see: \cite{2009A&A...499..923K}, \cite{2010ITPS...38.2249V}, and \cite{2016IAUS..320..233H}. 

Note that for the code-to-code comparison of \cite{kaspraova2019}, shown in Figure~\ref{fig:radynflarixcomp}, \radyn\ and \flarix\ were made as similar as possible as concerns the physics included (e.g. the atoms treated in detail, the optically thin loss functions, the form of electron beam heating).

\subsection{PREFT}
Longcope and collaborators have developed a flare loop model which incorporates reconnection energy release by using thin flux tube \citep[TFT;][]{1981A&A....98..155S,2006ApJ...642.1177L} MHD equations. The axis of a thin flux tube evolves under a version of the ideal MHD equations expanded in small radius \citep{2010ApJ...718.1476G}.  The tube is assumed to lay entirely within an equilibrium current sheet separating layers of magnetic flux with equal field strength but differing in direction by a shear angle $\Delta\theta$.  The magnetic pressure from the external layers maintains the field strength of the tube, but otherwise exerts no force on the tube's axis.  The tube evolves without reconnection under its own magnetic tension, and field-aligned pressure and viscosity.  Energy transport is assumed to occur through thermal conduction, limited to the level of free-streaming electrons.   The tube is initialised at the instant after a localised reconnection process within the current sheet has linked sections of equilibrium tubes from opposite sides of the current sheet.  No further reconnection occurs, and any heating from the initialising event is neglected.    In its subsequent evolution, the tube retracts under magnetic tension releasing magnetic energy and converting it to bulk kinetic energy in flows which include a component parallel to the tube.  The collision between the parallel components generates a pair of propagating slow magnetosonic shocks, which resemble gas dynamic shocks as they must in the parallel limit.  Absent thermal conduction, this evolution matches the classic models of Petschek reconnection including a guide field \citep[i.e. component reconnection][]{2009ApJ...690L..18L,2011ApJ...730...90G,2015ApJ...813..131L}.  Solutions of the TFT equations show that thermal conduction carries heat away from the shocks, drastically altering the temperature and density of the post-flare plasma \citep[][]{2011ApJ...740...73L,2015ApJ...813..131L,2016ApJ...833..211L}.

\cite{2010ApJ...718.1476G} reported the first numerical implementation of the coronal TFT equations in a code called \deft\ (Dynamical Evolution of a Flux Tube).  A later implementation, called \preft\ (Post-Reconnection Evolution of a Flux Tube; written in IDL), included optically thin radiative losses and a simplified chromosphere at each end of the reconnected flux tube, capable of reproducing chromospheric evaporation \citep{2015ApJ...813..131L}.  Both versions have been adapted to include current sheets terminating in Y-points \citep[][]{2011ApJ...730...90G,2020ApJ...894..148U}, although simulations are often done with a simpler uniform current sheet. Interactions with the current sheet plasma in the form of an aerodynamic drag force was later added \citep{2021ApJ...923..248U}, and recent experiments impeded this drag force to investigate the role of MHD turbulence in prolonging heating into the flare gradual phase (Dr. William Ashfield, \textsl{private communication}, 2022). The chromosphere in \preft\ is typically set to be isothermal ($T \sim 10$~kK) and gravitationally stratified according to some density scale height.

%%%%%%%%%%%%%%%%%%%%%%%%%%%%%%%%%%%%%%%%%%%%%
%% HOW TO FORWARD MODEL OBSERVABLES
%%%%%%%%%%%%%%%%%%%%%%%%%%%%%%%%%%%%%%%%%%%%%

\section{Forward Modelling IRIS Observables}\label{sec:synthesise}
Armed with the flaring atmospheres from the dynamic loop models we must then synthesise the emission that we predict IRIS would observe. This is true even in the case of \radyn\ and \flarix\, where the radiation transfer of certain species is solved alongside the hydrodynamics, since the spectral lines observed by IRIS are not typically included in those solutions. This generally happens in one of two ways: (1) via an optically thin `coronal' assumption, using data from an atomic database such as CHIANTI alongside instantaneous properties of the flare atmospheres (e.g., emission measures, temperatures, and velocities); (2) via detailed NLTE radiation transfer calculations using snapshots from the dynamic simulations as input to post-processing code. Within the context of IRIS, the former is typically done for the Fe~\textsc{xxi}, Si~\textsc{iv}, and other optically thin transition region lines, and the latter for Mg~\textsc{ii}, C~\textsc{ii}, and other optically thick transitions or lines such as \ion{O}{i} 1355.6~\AA\ for which certain processes such as charge exchange require a radiation transport approach. Each approach has drawbacks and advantages.

In the optically thin line synthesis scenario the usual approach is to generate the contribution functions $G(n_{e},T)$ (via a resource such as CHIANTI) that encapsulate the atomic physics processes that lead to the population and subsequent de-excitation of the transition in question. Normally this assumes ionisation equilibrium so that they peak at the equilibrium formation temperature of the ion in question. This may or not be a valid assumption in some scenarios. Codes like \hydrad\ that track the non-equilibrium ionisation of minority species can instead employ more realistic ionisation stratification. From these functions the emissivity in each grid cell can be calculated using the local plasma conditions within that cell:

\begin{equation}
j_{\lambda,z} = A_{b}G(n_{e},T)n_{e}(z)n_{H}(z), 
\end{equation}

\noindent where $A_{b}$ is the abundance of the species, $n_{e}(z)$ is the electron density at height $z$, and $n_{H}$, is the hydrogen density. The intensity in each grid cell is then $I_{\lambda,z} = j_{\lambda,z}\delta z$, for a grid cell width $\delta z$. Once the intensity is known the spectral lines can then broadened by the instrumental profile, by the Gaussian thermal width, and by any assumed non-thermal width (e.g. due to microturbulence). Non-thermal broadening is the catch all term for any width in excess of the quadrature sum of the thermal and instrumental widths, the source of which is an active area of study. This could be due to local turbulence, unresolved flows, or something more exotic. When modelling it is usually not included without some \textsl{apriori} information (or guess!) about what it should be.  Often some scheme is used to sum the intensity in each cell through height to provide the total emergent intensity. For example, this can be as basic as integrating through the full loop, or we can isolate certain locations such as the footpoints to integrate over. Other techniques attempt to take into account instrument properties, for example we can assume a semi-circular loop at disk centre, orientated perpendicular to the Sun's surface, and project each loop position onto an artificial pixel of some size inspired by the instrument we are trying to compare to \citep[e.g.][which is the norm for \hydrad\ simulations]{2011ApJS..194...26B}.  

When modelling optically thick lines by inputting snapshots from the flare simulations to radiation transport codes then it is usually the case that more advanced physics can be included in the solution that present in the original simulations, but at the expense of the dynamics. That is, non-equilbrium effects are often neglected and the statistical equilibrium population equations are solved instead. This can be mitigated somewhat by using in each atmospheric snapshot the electron density computed from a non-equilibrium solution, and in fact \cite{2019ApJ...885..119K} demonstrated using the minority species version of \radyn\ (\msradyn) that non-equilibrium effects can be, mostly, safely neglected when considering Mg~\textsc{ii} even in flares \citep[as was previously shown to be the case in the quiet Sun,][]{2013ApJ...772...89L}. In that instance, the inclusion of more advanced physics afforded by codes such as \rhpar\ \citep{2001ApJ...557..389U,2015A&A...574A...3P}, namely partial frequency redistribution (PRD), affected the line synthesis a great deal more than non-equilibrium effects. Post-processing flare snapshots through radiation transport codes typically involves providing the atmospheric stratification (e.g. height / column mass scale, temperature, electron density, velocity, microturbulence, H level populations or mass density) and a series of model atoms to solve (which in some cases form the basis of the non-hydrogen background opacities). The model atoms contain data about the atomic levels, the various transitions (oscillator strengths, damping terms etc.,), thermal collisional excitation and ionisation rates, charge exchange rates, and the like. Thus, careful construction of the model atom is required to obtain a good result, including using the appropriate number of levels. There are three commonly used radiation transfer codes when processing flare atmospheres: \rh, \rhpar, \& \mali. Of those, \rh\ and \rhpar\ are probably the most commonly used, and are in fact very similar to each other, the latter being a parallelised version of the former, allowing multiple snapshots to solved simultaneously. This greatly speeds up the problem as each 1D snapshot can take up to some tens of minutes or longer to solve if a very large number of transitions are desired, especially if PRD is being used. 

Of the optically thick IRIS lines modelled in detail in flares, Mg~\textsc{ii} is the most common, likely because it offers exceptional (though still largely untapped) diagnostic potential. \cite{2019ApJ...883...57K} and \cite{2019ApJ...885..119K} explored the impact of various radiation transfer effects when considering Mg~\textsc{ii} in flares, partly in an attempt to determine if the large densities in flares meant that PRD could be neglected and the more computationally friendly complete frequency redistribution approach employed. Unfortunately it was the case that PRD was indeed required even in flares, but the hybrid angle averaged PRD approach of \cite{2012A&A...543A.109L} was found to adequately approximate the full angle-dependent PRD solution and save an order of magnitude in computational time. Further, \cite{2019ApJ...883...57K} demonstrated that using a model atom with too few levels (e.g. 3 levels-plus-continuum) produced different results than a larger model atom, in part due to the lack of cascades through upper levels present in larger model atoms \citep[e.g. the 10-level-plus-continuum atom of][]{2013ApJ...772...89L}. Aside from hydrogen, including other species in NLTE was found to not be required. A similar trade study has not yet been performed for other IRIS transitions in flares, though would be a worthwhile exercise. 

Note that a new radiation transfer code capable of including time-dependent effects (if sufficiently high-cadence snapshots are available) and processes such as PRD and overlapping transitions has recently been developed: \textsl{Lightweaver} \citep{2021ApJ...917...14O}. This exciting new resource has not yet been used to study IRIS observables (to my knowledge!) but should be employed in future efforts. 

%%%%%%%%%%%%%%%%%%%%%%%%%%%%%%%%%%%%%%%%%%%%%
%% MASS FLOWS
%%%%%%%%%%%%%%%%%%%%%%%%%%%%%%%%%%%%%%%%%%%%%

\section{`Electron Beam' Driven Mass Flows}\label{sec:massflows}
A major consequence of solar flare energy deposition is the driving of strong mass flows, which appear in spectral observations as Doppler shifted features either in the line core or as asymmetries in line wings. Both in early observations and modelling of flares \citep[e.g.][]{1985ApJ...289..414F,1985ApJ...289..425F,1985ApJ...289..434F,1989ApJ...346.1019F} a distinction was made between chromospheric evaporations that proceeded in a more `gentle' fashion and those that were `explosive'. Gentle evaporation  is subsonic, whereas explosive evaporation is supersonic, reaching 100s~km~s$^{-1}$, and is very impulsive in character, with a rapid rise to peak velocity. The overpressure and momentum balance from the explosive upflow scenario also produces downflows with speeds on the order a few $\times10$~km~s$^{-1}$ (chromospheric condensations) that are denser and propagate deeper into the chromosphere, appearing as redshifts in spectral lines. The Fisher studies determined that the energy flux delivered to the chromosphere and transition region was the deciding factor, with $F = 1\times10$~erg~s$^{-1}$~cm$^{-2}$ required to drive an explosive response. Those models used a fixed heating duration of $t=5$~s, and a fixed $E_{c} = 20$~keV low-energy cutoff for the electron beam, and we know also that the low-energy cutoff can also an important parameter in determining the character of upflows/downflows, \citep[e.g.][]{2015ApJ...808..177R}. Note also that electron beams are not the only means to drive explosive chromospheric evaporation, and that they can be driven by a strong heat flux, for example. 

As mentioned above, these upflows are produced by pressure gradients following flare heating, with momentum balance driving downflows. It is worth pointing out that the `dividing line' between upflows and downflows has been observationally identified in temperature space using \textsl{Hinode}/EIS data \citep[e.g.][]{2009ApJ...699..968M}. In the footpoint of a C class flare, ions forming $T> \sim1.5$~MK exhibited large blueshifts whereas  $T< \sim1.5$~MK exhibited redshifts (assuming ionisation equilibrium for their formation temperatures). Studying an X-class flare in which EIS observed several footpoint sources \cite{2022ApJ...936...85S} found similar results but with a range of flow reversal temperatures $T_{FR} \sim [1.35-1.82]$~MK. This flow reversal point is located within the flare transition region, and roughly identifies the location of a pressure imbalance, and therefore heating location. It is an important benchmark for models to meet, though only one study to my knowledge has used this to test the physical processes in models. \cite{2022ApJ...931...60A} modelled the flare observed by \cite{2009ApJ...699..968M} using \radyn, where they explored the effects of turbulent and non-local suppression of thermal conduction. Comparing the synthetic EIS profiles they found that suppression factors between 0.3 and 0.5 times that of the Spitzer values were most consistent with the observed flow reversal temperature, the magnitudes of upflows as function of temperature, and the non-thermal widths \citep[studied for the same flare by][]{2011ApJ...740...70M}. In their study \cite{2022ApJ...931...60A} included the turbulent mean free path in the line synthesis, acting as a source of non-thermal broadening. \cite{2022ApJ...936...85S} performed a very detailed observational study of the flows from many lines as observed by EIS and IRIS, as well as analysing the hard X-ray observations from RHESSI. Modelling those various flare footpoints, driven by non-thermal electron distributions inferred from the RHESSI observations as a function of time \citep[for example using the \radynarcade\ framework of][see below]{2020ApJ...900...18K} would be a worthwhile endeavour to explore the pattern of upflows versus downflows. The upcoming  Solar-C/EUV High-Throughput Spectroscopic Telescope \citep[EUVST;][]{2019SPIE11118E..07S} will provide capabilities comparable to IRIS but with a significantly broader temperature coverage, which should also be a valuable resource for such studies. 

I will not review the wealth of observational evidence and analysis of chromospheric evaporations and condensations here but refer the reader to recent reviews of EUV \citep{2015SoPh..290.3399M} and UV flare spectroscopy \citep{2021SoPh..296...84D}, and reference therein. Instead, in this section I highlight a few studies in which loop models of flares driven by typical non-thermal electron distributions were used to interpret signatures of mass motions in IRIS spectra, and in which IRIS spectra challenge the models.

\subsection{Long-Lived Flows}\label{sec:longlivedflows}
The high spatiotemporal resolution afforded by IRIS has led to a plethora of studies of the chromospheric evaporation and condensation processes in flares. An important spectral line that has been used extensively for studying hot flare plasma is \ion{Fe}{xxi} 1354.1~\AA, forming at around $T\sim11$~MK (in equilibrium conditions). This line exhibits large Doppler motions both in flare ribbon footpoints, and along the legs of flare loops, in excess of $v_{Dopp} = 100$~km~s$^{-1}$ and up to $v_{Dopp} \sim250-300$~km~s$^{-1}$ \citep[e.g.][]{2014ApJ...797L..14T,2015ApJ...811..139T,2015ApJ...799..218Y,2015ApJ...807L..22G,2015ApJ...803...84P,2016ApJ...816...89P,2015ApJ...805..167S}. At the same time the lines are initially extremely broadened, narrowing as the line returns to rest. Importantly, this line is entirely blueshifted within an IRIS spatial pixel \citep[e.g.][]{2015ApJ...799..218Y}. That is, it does not just have a blue wing asymmetry alongside a stationary component, as was generally the case with MK lines observed during flares with lower-resolution observatories, the implication being that IRIS is resolving the flare footpoint source (if the filling factor is $\sim 1$), or that there was only one source of footpoint emission within that pixel (if the filling factor $< 1$).  

To set these data in context I include a brief aside to discuss pre-IRIS observations of spectral lines produced in plasmas with a temperature in excess of several MK, though encourage the reader to see \cite{2015SoPh..290.3399M} for a fuller discussion. Blueshifts of up to a few hundred km~s$^{-1}$ from lines at $>\sim8$ MK temperatures in flares were initially observed to possess a dominant stationary component plus a blueshifted component, for example by \cite{1982SoPh...78..107A}, \cite{1983SoPh...86...67A}, \cite{1986ApJ...309..435M}, and \cite{1989ApJ...344..991F} using SMM data. Further flare observations from \textsl{Yohkoh's} Bragg Crystal Spectrometer \citep[BCS;][]{1991SoPh..136...89C} found similarly asymmetric profiles. Such observations were contrary to the expectations of fully blueshifted lines based on numerical models in which the whole flare was a monolithic loop\footnote{Note that when the codes discussed in detail in this review are used as single loops it is not believed that they represent the entire flare volume, rather that they are some portion of it.} \citep[see discussions in ][]{1998ApJ...500..492H,2005ApJ...629.1150D}. These data lacked spatial information, so the total emission was the sum of all sources in the field of view.  However, even with instruments that provided spatial information, for example SOHO's Coronal Diagnostic Explorer \citep[CDS, with $4-5^{\prime\prime}$ resolution;][]{1995SoPh..162..233H} or \textsl{Hinode's} EUV Imaging Spectrograph \citep[EIS, with $\sim3^{\prime\prime}$ resolution;][]{2007SoPh..243...19C}, high-temperature lines still exhibited a stationary component alongside a blueshifted component. The dominance of each component varied, so that in some of those observations the blueshifted component was brighter than the stationary. Also, sometimes instead of being stationary, one component was actually just less blueshifted (still suggesting contributions from unresolved features). Some examples of studies that found multi-component lines include: \cite{2006A&A...455.1123T,2006ApJ...638L.117M,2009ApJ...699..968M,2011ApJ...727...98L,2011A&A...532A..27G,2010ApJ...719..213W,2013ApJ...766..127Y,2013ApJ...767...55D}. There were, however, some flares in which fully blueshifted high-temperature lines was observed, \citep[e.g.][]{2006SoPh..234...95D,2013ApJ...762..133B,2015ApJ...813...32D}. So, prior to IRIS there was not a consistent picture. Now with the consistent presence of fully blueshifted \ion{Fe}{xxi} 1354.1~\AA\ from IRIS we can more confidently isolate the hot flare footpoint emission to compare against predictions of mass flows in our numerical models.

A superposed epoch analysis of IRIS \ion{Fe}{xxi} 1354.1~\AA\ Doppler motions during an X class flare revealed a remarkably uniform behaviour within each footpoint \citep[][]{2015ApJ...807L..22G}. Along the flare ribbon each footpoint initially showed \ion{Fe}{xxi} $v_{Dopp} \sim250$~km~s$^{-1}$, with very little scatter, followed by a smooth decay in time back towards rest. Again with very little scatter, it took around $10$~minutes for each source to return to rest (similar timescales have been seen in other flares). For various reasons (e.g. ribbon propagation timescales, rise times of UV or optical emission, duration of hard X-ray spikes) it is generally assumed that energy injection into each footpoint is more on the order of seconds to tens of seconds. In models that employ those timescales the atmospheres undergo rapid global cooling, from flare temperatures towards quiescent temperatures, following cessation of energy injection and flows are quenched due to the collapse in chromospheric/transition region overpressure that drives the upflow of material. What then sustains these long lived upflows? 

Similarly, in a number of flares the transition region \ion{Si}{iv} resonance lines have been observed to exhibit redshifts lasting many minutes, in contrast to relatively shortlived chromospheric redshifts. Lifetimes of \ion{Si}{iv} redshifts range from a few tens of seconds, to minutes, or even tens of minutes, seen in both high and low cadence observations \citep[e.g.][]{2015ApJ...810...45B,2015ApJ...805..167S,2016ApJ...829...35W,2016ApJ...832...65Z,2015ApJ...811....7L,2017ApJ...848..118L,2019ApJ...879...30L,2020ApJ...896..154Y,2022ApJ...926...23L,2022ApJ...926..164A}. Pervasive net-redshifts, from $\sim1-15$~km~s$^{-1}$, with spatial fine structure, are not uncommon in the non-flaring mid-lower transition region, \citep[e.g.][]{1993ApJ...402..741H,1993ApJ...408..735B,1998ApJS..114..151C,2018A&A...614A.110Z}, and this may muddy the waters of identifying temporal signatures of condensations in transition region lines. For example several authors show observations of \ion{Si}{iv} Doppler motions in which there is a consistent $v_{Dopp}\sim10$~km~s$^{-1}$ even adjacent to the main flaring region, on top of which briefer (a few seconds to tens of seconds) bursts of redshifts occur co-spatial with flare sources, with magnitudes $v_{Dopp}\sim10-80$~km~s$^{-1}$ \citep[][]{2019ApJ...879...30L,2020ApJ...896..154Y,2022ApJ...926..164A}. Some do show smoother, less bursty Doppler shift lightcurves \citep[e.g.][]{2016ApJ...829...35W}. Flare induced Doppler motions typically appeared as asymmetries in the red wing, with some instances exhibiting a fully shifted profile or even a transition from fully shifted to asymmetric \citep[see discussions in][]{2019ApJ...879...30L,2020ApJ...896..154Y}.  Quasi-periodic pulsations (QPPs) have been identified in the \ion{Si}{iv} lines, with periods on the order $t\sim32-42$~s \citep{2016ApJ...832...65Z}, an interpretation of which could be so-called bursty reconnection with repeated energy injection (though the origin of QPPs is still a source of healthy debate). Other periodicities have been interpreted as being the result of current sheet dynamics \citep[][]{2015ApJ...810....4B}. Regarding the chromospheric redshifts, pre-IRIS observations suggested lifetimes of $2-3$~minutes for the decay of \ion{H}{i}~$\alpha$ redshifts \citep[e.g. the well-known results of][]{1984SoPh...93..105I,1995SoPh..158...81D}, but high spatiotemporal (so that effects of unresolved flows are reduced) observations of the \ion{Mg}{ii} lines from IRIS have suggested $\sim30-60$~s decay time \citep[e.g][]{2015ApJ...807L..22G,2015A&A...582A..50K}. In the IRIS \ion{Mg}{ii} observations, the redshifts rapidly fall on those short timescales, but there can be a residual low-magnitude redshift for a few minutes. The initial rapid decrease, though, is comparable to the theoretical predictions of \cite{1989ApJ...346.1019F}. A conceptual understanding of the briefer bursts of strong transition region redshifts as being due to the classical condensation picture is not difficult, but the cases with more sustained (over many minutes or tens of minutes), elevated, redshifted emission is rather more difficult to explain; especially as the chromospheric counterpart of the condensation does not tend to show such long periods. 

In this section I discuss some attempts to address the long duration upflows and downflows identified in IRIS observations with loop modelling, that focussed on the \ion{Fe}{xxi} and \ion{Si}{iv} lines. 

\subsubsection{Single Loop or Arcade Modelling}
\begin{figure}[h]
\begin{center}
\includegraphics[width=\textwidth]{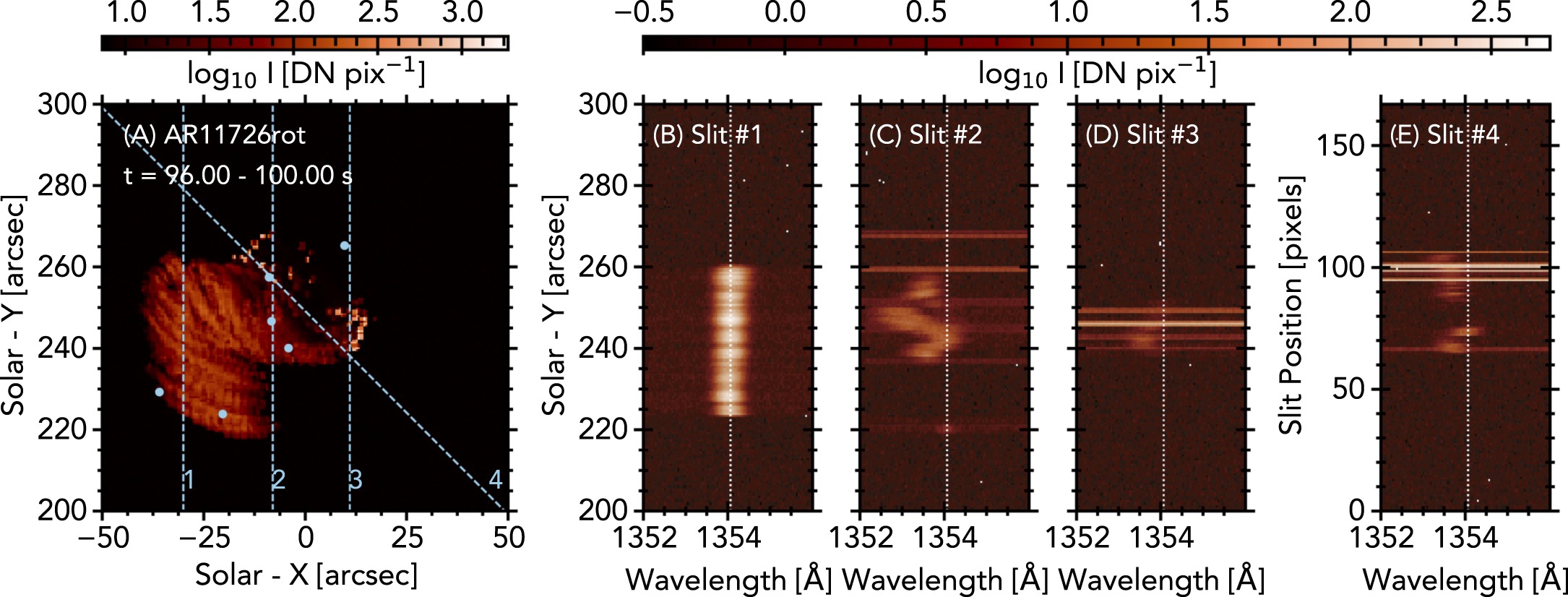}% This is a *.eps file
\end{center}
\caption{An example of \radynarcade\ modelling of \ion{Fe}{xxi} 1354.1~\AA, from \cite{2020ApJ...900...18K}. Panel \textbf{(A)} shows a map of the emission integrated over the \ion{Fe}{xxi} line, with bright newly activated footpoints, and dense loops. Dashed lines are artificial slits, along which the spectra are shown in panels \textbf{(B-E)}, where Doppler shifts and broadening is present. Bright horizontal strips are intense continuum enhancements at loop footpoints. \copyright AAS. Reproduced with permission.}\label{fig:arcademodel}
\end{figure}

\begin{figure}[h]
\begin{center}
\includegraphics[width=\textwidth]{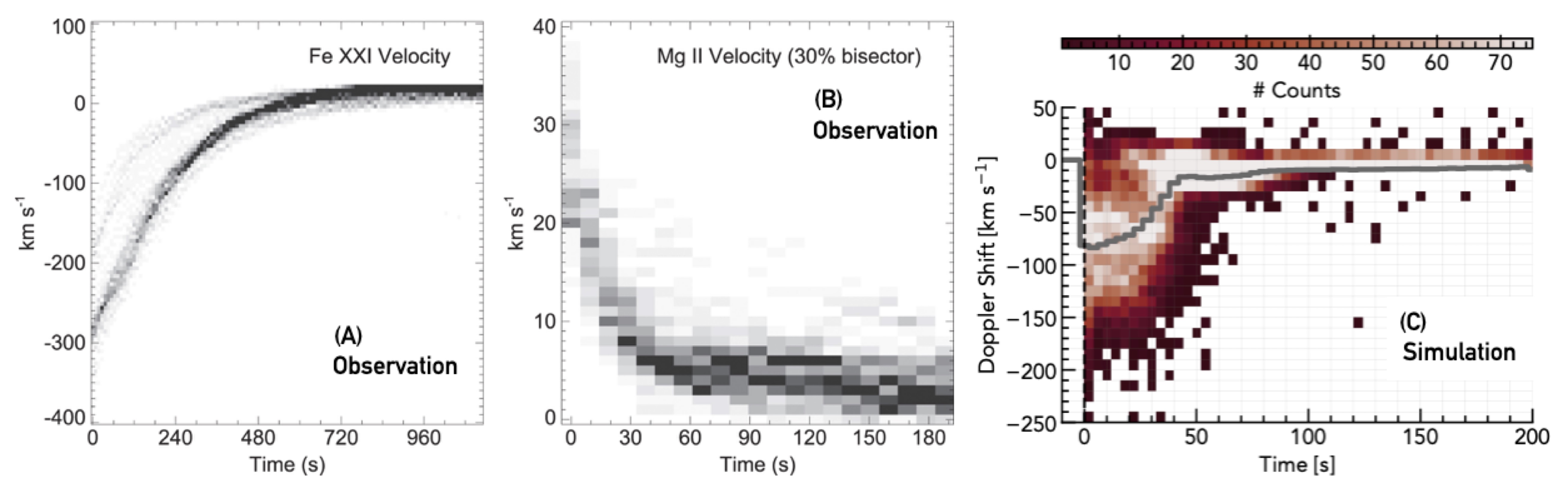}% This is a *.eps file
\end{center}
\caption{Comparing an observed superposed analysis of mass flows to the modelled upflows. Panel \textbf{(A)} shows the \ion{Fe}{xxi} 1354.1~\AA\ Doppler shifts obtained from fitting Gaussians to IRIS observations of the 2014-September-10th X-class flare, and panel \textbf{(B)} shows the \ion{Mg}{ii} subordinate line Doppler shifts from that same flare, obtained via a bisector analysis (negative velocities are blueshifts, positive are redshifts). Both are adapted from \cite{2015ApJ...807L..22G}. Panel \textbf{(C)} shows the synthetic \ion{Fe}{xxi} 1354.1~\AA\ line Doppler shifts from the \radynarcade\ model of \cite{2020ApJ...900...18K}, where the grey line represents the mean. Clearly the modelled upflows subside an order of magnitude too quickly. \copyright AAS. Reproduced with permission.}\label{fig:superposedfig}
\end{figure}

In an effort to facilitate a more realistic model-data comparison of optically thin flare emission, \cite{2020ApJ...900...18K} produced a synthetic flare arcade model that used an observed active region magnetic skeleton and \radyn\ field-aligned models. This model takes into account the superposition of loops and geometric effects (e.g. loop inclination, viewing angles) so that line-of-sight effects in the synthetic optically thin images and spectra are accounted for. A \radyn\ model was grafted onto magnetic loops extrapolated from a non-flaring active region \citep{2018arXiv180700763A}, and were set off in sequence to mimic ribbon propagation (5 loops every 3 seconds). Within each voxel of the 3D space the differential emission measures (DEM; a measure of how much material is present within a temperature bin) and \ion{Fe}{xxi} 1354.1~\AA\ spectra were synthesised, and, from the former, observables from the Solar Dynamics Observatory's Atmospheric Imaging Assembly \citep[SDO/AIA;][]{2012SoPh..275...17L} were synthesised. These were projected onto a 2D observational plane, with multiple voxels projected into a single pixel, so that superposition along the line-of-sight was included. This flare arcade model reproduced many aspects of \ion{Fe}{xxi} Doppler shifts such as the magnitude of the blueshifts, the narrowing as they approached rest, and the localisation of the blueshifts to hot footpoints and lower legs of the loops. Figure~\ref{fig:arcademodel} shows some synthetic \ion{Fe}{xxi} spectra and a map of the arcade in this model. However, this model significantly under-predicted the decay time. Constructing a superposed Doppler flow similar to \cite{2015ApJ...807L..22G} showed only a $t\sim50$~s decay time of footpoint Doppler shifts, more than an order of magnitude too fast. Each loop in the arcade was from the same \radyn\ simulation, with $t\sim25$~s injection time, but the projected velocity differed depending on the loop geometry. Figure~\ref{fig:superposedfig} illustrates the differences in decay time between the observations of \cite{2015ApJ...807L..22G} and the flare modelling. 

This result was perhaps not unexpected. It is generally the case that flows subside not long after the cessation of energy injection, which alongside rapid global cooling removes the pressure imbalance in the absence of heating, quashing the upflows. \cite{2018ApJ...856..149R} contains some detailed examples of this, where upflows in either a single model with bursty injection, or a multi-threaded model with simultaneous individual heating events, disappeared shortly after the electron beams were switched off. As well as unexplained lengthy upflows we have a mix of short and long duration condensations as discussed above, which are also frustratingly hard to explain. The \cite{1989ApJ...346.1019F} models of condensations predict that the downflows last $t\sim30-60$~s, almost regardless of particle injection timescales. Once the chromosphere has been shocked out of equilibrium, the condensation propagates deeper, but accrues mass as it does so, and decelerates. Even if energy release is continuous over an extended time it seems hard to drive longer lived downflows as the chromosphere reaches a new equilibrium and is no longer shocked. The chromospheric timescales predicted by models are not as incongruous with observations as the upflows into the corona. Doppler shifts observed in \ion{Mg}{ii} and other chromsopheric lines have lifetimes not too dissimilar from the modelled $30-60$~s \citep[e.g.][]{2015ApJ...807L..22G,2020ApJ...895....6G}, shown in the middle panel of Figure~\ref{fig:superposedfig}.

One means to obtain long duration upflows of \ion{Fe}{xxi} 1354.1~\AA\ is simply to bombard the atmosphere with an electron beam for an extended period of time. To reproduce the $\sim1000$~s of upflows observed in the 2014-Oct-27th X class flare, \cite{2015ApJ...803...84P} heated a single monolithic \hydrad\ loop for $1000$s of seconds. They synthesised \ion{Fe}{xxi}, degraded to IRIS resolution, and were able to obtain Doppler shifts with a very similar decay profile (though it was necessary to divide the energy flux inferred from RHESSI observations by a factor of 10 to avoid excessively high temperatures and densities compared to their DEM analysis). Curiously, running with and without non-equilibrium effects turned on seemed to suggest an observed Doppler flow pattern somewhat more consistent with the loop being in equilibrium. This may indicate that the electron density during the actual flare was larger than the \hydrad\ model. \cite{2015ApJ...803...84P} also tried asymmetric energy injection, where the electrons were injected somewhere along the loop, rather than the apex. The maximum of temperature and velocity occurred higher in the loop in that scenario, which may explain why the \ion{Fe}{xxi} footpoint emission has been observed to appear $1^{\prime\prime}$ from the enhanced FUV continuum, assumed to be the chromospheric footpoint \citep[e.g.][]{2015ApJ...799..218Y}. So, continuous heating can drive extended durations of Doppler shift. However, it is unlikely that electron beams are injected into one loop for 10 minutes or more. A number of researchers assume it is actually closer to being on the order of $<10-30$~s into each footpoint. This has been inferred from rapid ribbon propagation, indicating energy injection into a new loop, motion of hard X-ray sources, the duration of individual spikes in hard X-ray lightcurves, and the rise time of UV and optical lightcurves from individual pixels. 

\subsubsection{Multi-threaded Modelling}
An alternative means to achieve long duration upflows, and to address similarly long lived transition region downflows is multi-threaded modelling in which each IRIS pixel is assumed to be comprised of many strands embedded within that volume, with one loop model representing just one of these strands. \cite{2016ApJ...829...35W} studied a B2 class flare, noting that whereas \ion{Mg}{ii} exhibited discrete bursts of intensity enhancements with associated redshifts, the \ion{Si}{iv} and \ion{C}{ii} resonance line redshifts were more systematic, not returning to rest for many minutes. Their lightcurves were also sustained but showed more structure, with a few strong spikes. To explain these transition region observations \cite{2016ApJ...827..145R} used \hydrad\ loop models in a novel way. They ran 37 electron beam driven simulations, with the spectral index and low-energy cutoff inferred from RHESSI observations, and a range of energy fluxes spanning $F=10^{8-11}$~erg~s$^{-1}$~cm$^{-2}$, applied for $t=10$~s to each loop. From each loop, the \ion{Si}{iv} spectra were synthesised, applying Doppler shifts as appropriate. They then randomly sample emission from $N$ threads, activating randomly with an average rate of 1 per $r$ unit time, and with a power law in energy flux space guided by the FUV intensity distribution from observations of \cite{2016ApJ...829...35W}, dividing the total summed intensity by $N$. $N$ and $r$ are connected by the requirement that $N\times r> \tau$, for $\tau$ some observed duration, for example the duration of redshifts.  Within each individual strand, weak energy deposition ($<\sim10^{9}$~erg~s$^{-1}$~cm$^{-2}$) only produced blueshifts, due to gentle evaporation, consistent with the modelling of \citealt{2014Sci...346B.315T}. Stronger energy deposition quickly produced redshifts and high intensities. 

Through trial and error to ensure smooth and long-lived \ion{Si}{iv} Doppler shift lightcurves, \cite{2016ApJ...827..145R} determined that to reproduce the \cite{2016ApJ...829...35W} observations, $N\ge60$ threads were need per IRIS pixel, activating with a rate $r \le 10$~s, and with a minimum flux of $F=3\times10^{9}$~erg~s$^{-1}$~cm$^{-2}$. Burstier lightcurves resulted from decreasing the activation rate (increasing $r$), or allowing a smaller minimum energy so that large excursions resulting from the strongest threads were more prominent (recall that some observations of larger flares showed quite bursty \ion{Si}{iv} Doppler motions). Comparing the observed and modelled slopes of the emission measure distributions (EMD) can help further constrain $N$ and $r$. A smaller $r$ forces a higher $N$, the consequence of which means there are many more cooling loops at any time, boosting the low temperature end, and thus the slope, of the EMD at those temperatures. 

Building upon this framework, \cite{2018ApJ...856..149R} performed similar modelling that could simultaneously address the long lived \ion{Fe}{xxi} upflows and \ion{Si}{iv} downflows. They again kept $\delta = 5$ and $E_{c} = 15$~keV of the non-thermal electron distribution fixed, but this time allowed the duration of energy injection onto each thread to vary. A triangular pulse with equal rise and fall times, with total durations ranging $t_{dur} = [1-1000]$~s, in increments of 0.1 in log space was used, with peak energy fluxes ranging $F_{peak} = [10^{8} - 10^{11}]$~erg~s$^{-1}$~cm$^{-2}$ (providing 341 simulations total). Both \ion{Si}{iv} and \ion{Fe}{xxi} emission was synthesised from each loop model under optically thin conditions. Comparing two of their single loop models, \cite{2018ApJ...856..149R} first demonstrated that with longer energy deposition timescales, the upflows persist, but that downflows diminish even before the energy deposition ceases, confirming earlier results from Fisher's models. Once the chromosphere has been shocked and produces a condensation, it is difficult to re-shock just with continuous energy deposition. Further, the weaker simulations did not produce sufficiently hot, sufficiently dense loops to emit strongly in \ion{Fe}{xxi} (for intermediate energy fluxes non-equilibrium effects could lead to a delay in the formation \ion{Fe}{xxi}). Comparisons of the peak densities, temperatures, and mass flows produced by the grid of individual loops demonstrated that in addition to the energy flux, and low-energy cutoff, the duration of energy deposition likely plays a role in determining if evaporation proceeds explosively or not.  

Performing the multi-threaded modelling with various setups (variously fixing or changing values of $N, r, F_{min}, \alpha$, where $\alpha$ describes the slope of the energy flux distribution) using this \hydrad\ grid, \cite{2018ApJ...856..149R} found the following general conclusions. Longer heating durations produce smoother lightcurves, as loops heated by shorter durations cool too fast. For a mix of heating durations, a median of $t_{dur}=[50-100]$~s does a better job of reproducing the combination of long lived up- and downflows. More sustained past heating $t_{dur}> \sim100$~s results in the initial set of loops dominating the signal, but their downflows still subside after $\sim1$~minute so that a persistent redshift isn't seen. Much shorter durations make it difficult to produce detectable \ion{Fe}{xxi} emission. Increasing the number of loops produces smoother lightcurves, which for the redshifts is due to the dominance of newly activated loops. 

\begin{figure}[h]
\begin{center}
\includegraphics[width=\textwidth]{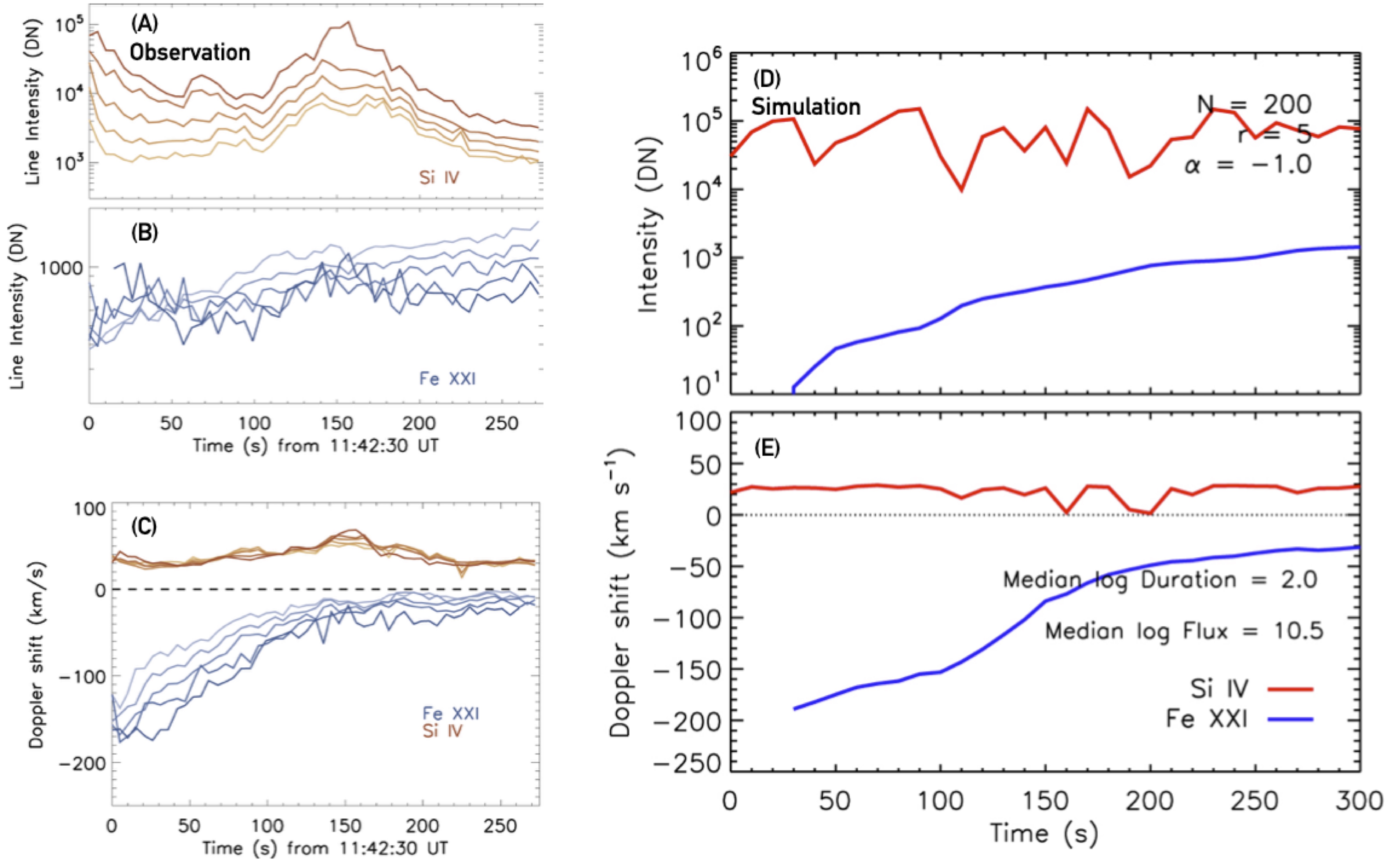}% This is a *.eps file
\end{center}
\caption{Explaining the observed long-duration flows with multi-threaded modelling. Panels \textbf{A-C} show observations of the M class 2015-March-12 solar flare, where from top to bottom we see the \ion{Si}{iv} 1403~\AA\ intensity (yellow/brown colour scale), the \ion{Fe}{xxi} 1354.1~\AA\ line intensities (blue colour scale), and the Doppler shift of each line(negative velocities are blueshifts, positive are redshift). The multiple lines shown for each quantity represent five different pixels in the flare ribbon. Panels \textbf{D-E} show the equivalent properties, but for a multi-threaded \hydrad\ simulation, with $N = 200$ threads per IRIS pixel. In those panels red shows \ion{Si}{iv} and blue shows \ion{Fe}{xxi}. This model does a good job at capturing the duration of flows but does not exhibit the decrease in \ion{Si}{iv} intensity towards the latter stage of the flare. Figure adapted from \cite{2018ApJ...856..149R}. \copyright AAS. Reproduced with permission.}\label{fig:multithreaded}
\end{figure}

Applying this model to estimate the number of strands per IRIS pixel in an observed flare, the M class event on 2015-March-12, \cite{2018ApJ...856..149R} find they can largely reproduce aspects of the observations with $r = [3,5]$~s, and heating durations between $t_{dur} = [30-300]$~s. However, they note that the observed \ion{Si}{iv} intensities decrease with time, possibly indicating that the maximum energy flux in the distribution of threads decreases over time. This comparison is shown in Figure~\ref{fig:multithreaded}, with the observations on the left, and the $N=200$, $r = 5$~s multi-threaded \hydrad\ model on the right. Comparing to Figure~\ref{fig:superposedfig}, the multi-threaded approach clearly does a better job at reproducing the extended Doppler shifts. 

To tackle the chromospheric predictions in their multi-threaded modelling \cite{2019ApJ...871...18R} synthesised \ion{O}{i} 1355.6~\AA\ and \ion{Mg}{ii} k line spectra via \rhpar\ with \hydrad\ flare atmospheres as input, after first modifying the treatment of the chromosphere in \hydrad. Before \cite{2019ApJ...871...18R} the H ion fraction, and therefore estimates of radiative losses from the \cite{2012A&A...539A..39C} lookup tables, in the \hydrad\ chromospheres were based on an LTE treatment that used the local temperature and electron density, assuming collisional rates dominated. To improve this, and to obtain better estimates of H ionisation stratification, electron density stratification, and radiative losses, they turned to the approach of \cite{2007A&A...473..625L}, who followed the work of \cite{1999MsT..........1S}. In this model the radiative and collisional rates for each transition of H were considered in order to obtain level populations for five levels of H plus the proton density. The collisional rate data were taken from standard sources. Obtaining the radiative rates without calculating the radiation field by solving the full NLTE radiation transfer problem required estimating the radiation field stratification above some critical height (below which the atmosphere could be assumed to be LTE), varying as a function of column mass above this height. Ultimately, the brightness temperatures were obtained as a function of height, and used to calculate the radiative rates. \cite{2019ApJ...871...18R} follow the \cite{1999MsT..........1S} results, but make some important modifications to account for the fact that the radiation field at the top of the chromosphere differs in flares. They note the relation between electron density and line intensity, using this to vary the brightness temperature at the top of the atmosphere as a function of electron density, with a grid of \radyn\ models serving as a guide. 

Armed with this new model chromosphere, the parameter space of \cite{2016ApJ...827..145R} was re-run to model the observations of \cite{2016ApJ...829...35W}, and multi-threaded models of the chromospheric lines calculated via processing \hydrad\ flare atmospheres through \rhpar\ and combining the emission in the same manner as \cite{2016ApJ...827..145R}. \cite{2019ApJ...871...18R} argue that the single loop model cannot explain the \ion{O}{i} emission, since it is strongly redshifted in the model but largely stationary in the observations. It was also far too bright in the models. Various setups of the multi-threaded approach managed to produce \ion{O}{i} Doppler motions consistent with \cite{2016ApJ...829...35W} observations, but the ratio of \ion{O}{i} to \ion{C}{i} was not consistent (\ion{O}{i} was too bright). The \ion{Mg}{ii} results were also consistent with the observations, showing bursty redshifts lasting throughout the heating phase. Something not addressed by \cite{2019ApJ...871...18R} is if we might really expect the chromospheric emission in a multi-threaded model to freely escape without radiation transfer effects between each closely space thread, which may confuse the picture of the very optically thick \ion{Mg}{ii} lines in this model (though \ion{O}{i} is very likely optically thin in flares). Single loop models likely suffer similar issues with 2D/3D radiation transfer, though often the assumption is that they are embedded in a ribbon-like structure that evolves similarly, rather than having many tens or hundreds random energisation events within the small volume of an IRIS pixel. A recent study using \radyn\ flare models and \textsl{Lightweaver} has demonstrated the importance of including 2D and 3D radiation transfer \citep{2022MNRAS.516.6066O}, though focussing on quiet Sun nearby a ribbon. A similar model that explores the effects \textsl{within} the ribbons or footpoints themselves (i.e. an inhomogenous ribbon) would be very interesting and worthwhile!

The application of multi-threaded modelling to the problem of long lived flows has been largely successful, with important implications if true, for example that individual threads may be smaller than $1/100$~arcseconds for $N\sim60$,  scaling inversely with $N$. While this model alleviates the demand of continuous energy injection into a single thread for many minutes, it still demands the injection for many minutes into an area the size of a single IRIS pixel (i.e. a single localised volume), and for up to $\sim2$~minutes on some threads. It remains to be seen if we actually have bombardment by energetic electrons into these areas for this long. Hard X-ray kernels, the proxy for non-thermal electrons,  propagate in time during the flare, so don't necessarily always hover in a single location for many minutes \citep[e.g.][among many studies]{2011SSRv..159...19F,2012ApJ...744...48C}, with the usual caveat that RHESSI's 1:10 dynamic range means that only the strongest sources are observed. Perhaps in the latter phase of multi-threaded modelling the upper range of the injected flux should be significantly reduced so that hard X-ray emission would be quite small, which is actually indicated by the pattern of \ion{Si}{iv} intensities in \cite{2018ApJ...856..149R}. It would also be an interesting comparison to study the lifetime of flows in events in which hard X-rays quickly vary location, to those in which the hard X-ray motions are relatively stable. Further (and perhaps related), chromospheric redshifts do not seem to always exhibit the long lived decays seen in the B class event of \cite{2016ApJ...829...35W}, and instead can show rapid decreases on the order of $[30-120]$~s. It would be interesting to apply the multi-threaded approach to a flare observed with longlived transition region and coronal flows, but shorter duration chromospheric condensations, to determine if some parameter set can be constrained. 

\subsubsection{Area Expansion in a Single Loop}
An important facet of loop models that is typically ignored when modelling solar flares is area expansion along the loop. In most flare loop modelling the loop is assumed to be semi-circular with uniform cross-sectional area. However, given that the magnetic field decreases with height from photosphere to corona, in order to conserve magnetic flux, the area of the loops should presumably expand with height. This could also help alleviate the stark model-data discrepancies of transition region spectral line intensities \citep[see the discussion in the detailed study of][]{2022ApJ...927..103R}. Including area expansion in our models is relatively straightforward, but the questions are by how much should the area expand, where should this expansion begin, and does this have a strong effect? 

The uncertainty here is not helped by the fact that it is observationally very tricky to identify the appropriate values to use. In fact, observations of both quiescent and flaring coronal loops do not seem to show significant expansion along their length, varying by only $\sim30~\%$ from midpoint to footpoint, \citep{2000SoPh..193...53K}. Even observations at the highest spatial resolution yet achieved show similar results \citep{2020ApJ...900..167K}. Nevertheless, it is important to gain an understanding of the effect on flare dynamics if we include area expansion. To that end, \cite{2022ApJ...933..106R} performed a systematic study of \hydrad\ flare simulations that included area expansion. Although they did not synthesise IRIS observables in this particular study, their results are applicable to the problem of long-lived flows and to the fact that flare models typically cool much faster than observations suggest \citep[see discussions in][]{2016ApJ...820...14Q,2018ApJ...865...67E,2022ApJ...931...60A}. Two scenarios were considered, one where expansion is limited to the transition region, and one where expansion occurs gradually and continuously through the loop, both implemented via a height varying factor $1/A(s)$ applied to the relevant hydrodynamic equations (where $s$ is the position along the loop). This factor, that describes the relative area expansion, was obtained by imposing a magnetic field stratification, and using the fact that the area expansion is proportional to the magnetic field decrease.

In their continuous-expansion experiments, the expansion factors, from footpoint to loop apex, tested were $A_{exp} = [1, 11, 43, 116]$. Each factor was used in an electron-beam driven flare simulation, with a deposition duration of $100$~s. The time taken to reach peak density was delayed with increasing area expansion, producing much longer cooling phases than typically seen in flare simulations. There is an extended period of time where the peak densities remain roughly constant and the temperatures decrease slowly via radiation, with draining only occurring after the temperature drops below $T = 100$~kK. Area expansion modified the $T\sim n^{2}$ relation so that the dynamics of the radiative cooling phase of the flares were very different than a loop with uniform cross-section. Upflows through the high temperature coronal loops persist well beyond the energy injection phase, in contrast to results discussed previously. Sound waves that result from sloshing of material during the flare gradual phase are also suppressed with increasing area expansion, and the magnitude of evaporative upflows reduced as the plasma encounters larger cross-sections. Sun-as-a-star irradiances synthesised from these models were reduced for loops with increasing area expansion but equal total volume, and the longer draining and cooling timescales results in sustained emission. Similar results were found for area expansion localised near the transition region, but with smaller changes to the timescales compared to the continuous-expansion case when the expansion occurs closer to the flare footpoint in the transition region. Sounds waves were also less suppressed in this scenario.

The assumption of a semi-circular loop was also interrogated, with a modification made to the gravitational acceleration term parallel to the loop to make the loops more elliptical. This has a seemingly minor effect, mostly on the draining timescales due to slightly weaker gravitational acceleration.  

It is not yet known what the appropriate values of area expansion to use are, but \cite{2022ApJ...933..106R} has convincingly demonstrated that this factor should not be ignored, particularly for the gradual phase of each footpoint. Indeed, this may negate the requirement for continuous energisation of many threads within a single IRIS pixel in order to maintain long-lived flows. Hopefully further exploration of these impacts will shed light on the appropriate values to use. 

\subsection{Satellite Component Redshifts}\label{sec:downflows}
Redshifted, broadened, emission appearing in the wings of strong chromospheric lines has been observed for many decades, for example famously in H$\alpha$ \citep[e.g.][]{1984SoPh...93..105I} who found short-lived ($\sim30-40$~s) sources with H$~\alpha$ red-wing asymmetry with implied velocities of $40-100$~km~s$^{-1}$. Other important studies of chromospheric redshifts from ground based observations of H~$\alpha$, \ion{Ca}{ii} and Na D lines include \cite{1990ApJ...348..333C}, \cite{1992A&A...256..255F}, \cite{2002A&A...387..678F}, \cite{1997A&A...328..371F}, and \cite{1988ApJ...324..582Z}. \cite{1997A&A...328..371F} noted that redshifts tend to occur along the edge of propagating ribbons; that is, they occur at the feet of newly reconnected loops. Comparing the momentum in condensations observed in H~$\alpha$ to that of upflows from SMM data, \cite{1990ApJ...348..333C} found an order of magnitude consistency, bolstering the chromospheric evaporation models of \cite{1985ApJ...289..414F}, and ruling out some alternate suggestions for the origin of upflows. The observational studies of \cite{1992A&A...256..255F} and \cite{2002A&A...387..678F} found what appeared to be an extended spatial gradient in the condensation front, seemingly at odds with the models of \cite{1985ApJ...289..414F} and \cite{1989ApJ...346.1019F}, who predicted a narrow condensation. They do speculate, though, that if the condensations in fact originated from a smaller area (i.e. that flare footpoints are smaller than they were able to resolve) then the wing emission could originate from a high-lying condensation and not from the deeper region implied by the line core intensity. In that scenario, the condensation may still have a gradient, but this gradient would be sharp since the geometric extent of the feature is narrow. Indeed, most of the studies mentioned above, as well as others referenced by them, commented specifically on the need for improved spatial (sub-arcsecond) and temporal (the first few seconds of energy deposition)  resolution of chromospheric flare footpoints. As noted by \cite{2020ApJ...895....6G}, pre-IRIS, sufficiently resolved observations of the flare impulsive phase in the chromosphere were scarce (in part due to the difficulty of placing spectrograph slits in the correct place).

\begin{figure}[h]
\begin{center}
\includegraphics[width=\textwidth]{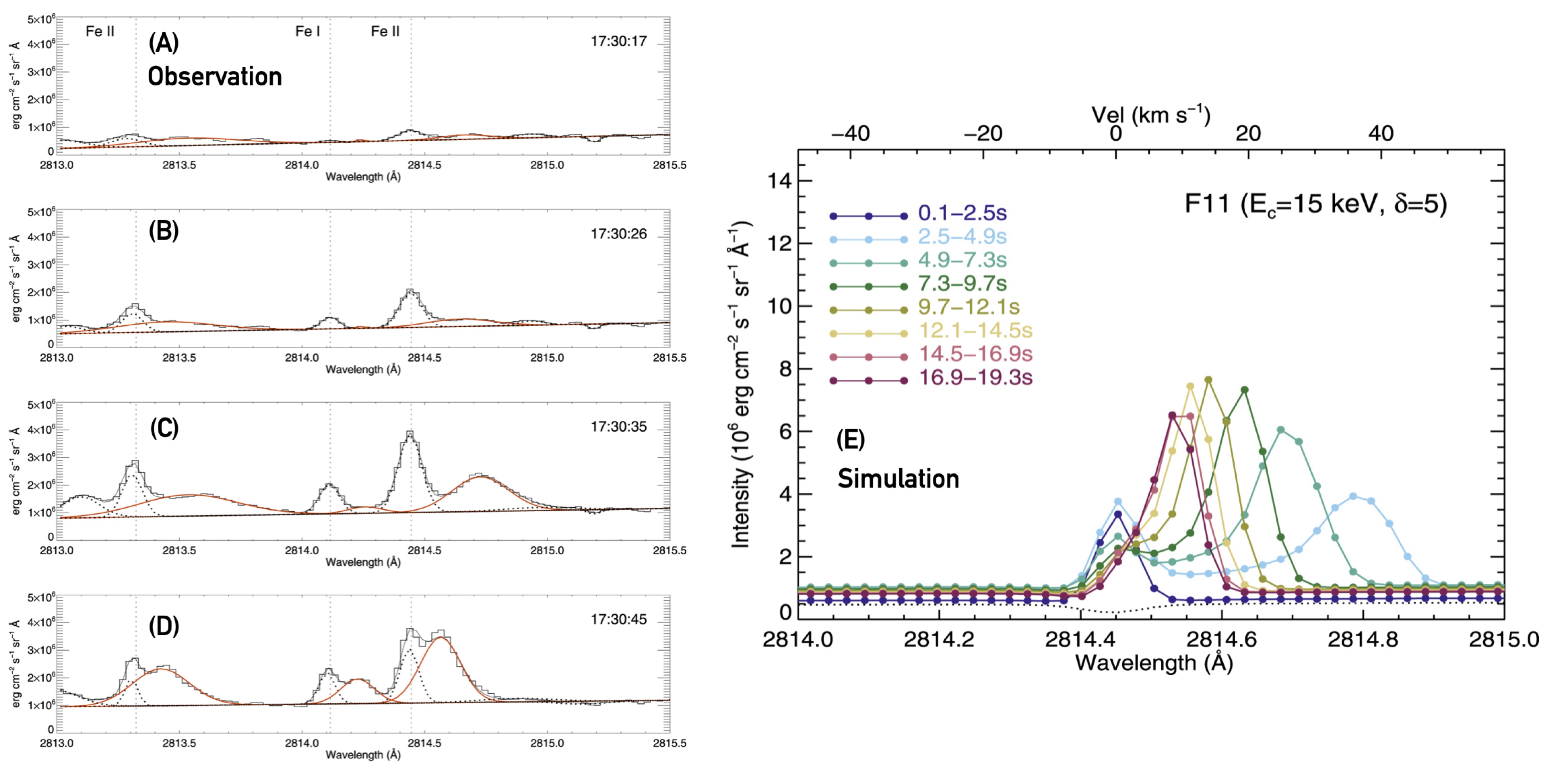}% This is a *.eps file
\end{center}
\caption{Modelling red wing asymmetries and satellite components. Panel \textbf{(A-D)} show a series of snapshots of the \ion{Fe}{ii} 2814.5~\AA\ line, observed by IRIS in the 2014-Sept-10 flare, where a stationary and satellite component are clear. The black solid histogram lines are the data, the black dotted line is the Gaussian fit to the stationary component, the red line is the Gaussian fit to the satellite component, and the thin grey line is the sum of the two components. Panel \textbf{(E)} shows \radyn\ modelling of the \ion{Fe}{ii} line from that flare, which was able to reproduce the satellite component, though with an overestimated intensity. Figure adapted from \cite{2020ApJ...895....6G}. \copyright AAS. Reproduced with permission.}\label{fig:satellite}
\end{figure}

IRIS has now observed many example of chromospheric downflows in flares, especially in the Mg~\textsc{ii} NUV spectra, at sub-arcsecond resolution \citep[e.g.][]{2015A&A...582A..50K,2015SoPh..290.3525L,2015ApJ...807L..22G,2015ApJ...804...56R,2017ApJ...836...12K,2018ApJ...861...62P,2019ApJ...878L..15H,2020ApJ...895....6G}.  \cite{2017ApJ...836...12K} identified similar features in weaker, more narrow chromospheric lines observed by IRIS, where they could appear as distinct components that persisted only for the duration of one $75$~s raster. Modelling of the two brightest footpoints in that flare, using \radyn, revealed that a rather high energy flux of non-thermal electrons was required to be injected into each footpoint, but that the dwell time on each footpoint could be different. A short pulse ($\Delta t = 4$~s) and a longer pulse ($\Delta t = 8$~s) were necessary to produce consistent ratios of \ion{Fe}{ii} redshifted component intensity to line core intensity within each different footpoint. This bore similarities to the conclusions of \cite{2002A&A...387..678F} who suspected that the red-asymmetry of certain line wings could originate from a condensation at a greater altitude than the usual line formation height. 

Higher cadence ($\delta t = 9.4$~s) IRIS observations of the 2014-September-10th X-class flare revealed that in addition to spectra exhibiting redshifted cores and red-wing asymmetries in chromospheric species \citep{2015ApJ...807L..22G}, many spectra contained separate components with redshifts indicating downflows of the order $25-50$~km~s$^{-1}$. These components were at times sufficiently far from the strong, mostly stationary, components that they were dubbed `satellite' components  \citep{2020ApJ...895....6G}. These were most apparent in singly ionised and neutral transitions that produce spectral lines that were generally more narrow than the very strong resonance lines observed by IRIS. For example, they were seen in \ion{Mg}{ii} 2791.6~\AA, \ion{Fe}{i} 2714.11~\AA, \ion{Fe}{ii} 2813.3~\AA, \ion{Fe}{ii} 2814.45~\AA, \ion{C}{i} 1354.284~\AA, and \ion{Si}{ii} 1348.55~\AA. The lefthand side of Figure~\ref{fig:satellite} shows examples of these satellite redshift components for the \ion{Fe}{ii} 2814.45~\AA\ lines, where it can be seen that the satellite components were broader than the primary more intense `stationary' component, and were observed to migrate towards, and ultimately merge with, the primary component over a period of $\sim30$~s (that is, they decelerated). 

Using \textsl{Fermi}/GBM \citep[][]{2009ApJ...702..791M} hard X-ray data, \cite{2020ApJ...895....6G} performed data-driven modelling of the \ion{Fe}{ii} 2814.45~\AA\ line from that flare. The spectral properties of the non-thermal electron distribution were obtained ($\delta = 5$, $E_{c} = 15$~keV), along with the total instantaneous power carried as a function time, averaged over $10$~s time bins to be consistent with the IRIS data. Crucially, the energy flux (power / area) was estimated by carefully measuring the newly brightened area of IRIS SJI images at each time, with the rationale being that this revealed the locations into which the non-thermal electrons observed in Sun-as-a-star \textsl{Fermi} data were being injected at any snapshot. Using different thresholdings to define this area provided a range of energy flux densities on the order $10^{11-12}$~erg~s$^{-1}$~cm$^{-2}$. Finally, the duration of energy injection into each footpoint was estimated from the width of a half-Gaussian function fit to the rise time of the \ion{Fe}{ii} spectra \citep[similar to the approach of][]{2012ApJ...752..124Q}, revealing a dwell time\footnote{Here the assumption is that the sharp rise time from background to peak intensity is an indication of the characteristic heating duration. From the histogram of Gaussian widths a typical heating time (dwell time of the electron beam in this case) was obtained.} of $t_{inj}\sim20$~s. \radyn\ modelling was performed with these derived parameters as input, in which a prominent condensation rapidly formed, with downflowing speed of up to $50$~km~s$^{-1}$, an electron density in excess of $10^{14}$~cm$^{-3}$ and a width of only $\Delta z=30-40$~km. 

\begin{figure}[h]
\begin{center}
\includegraphics[width=\textwidth]{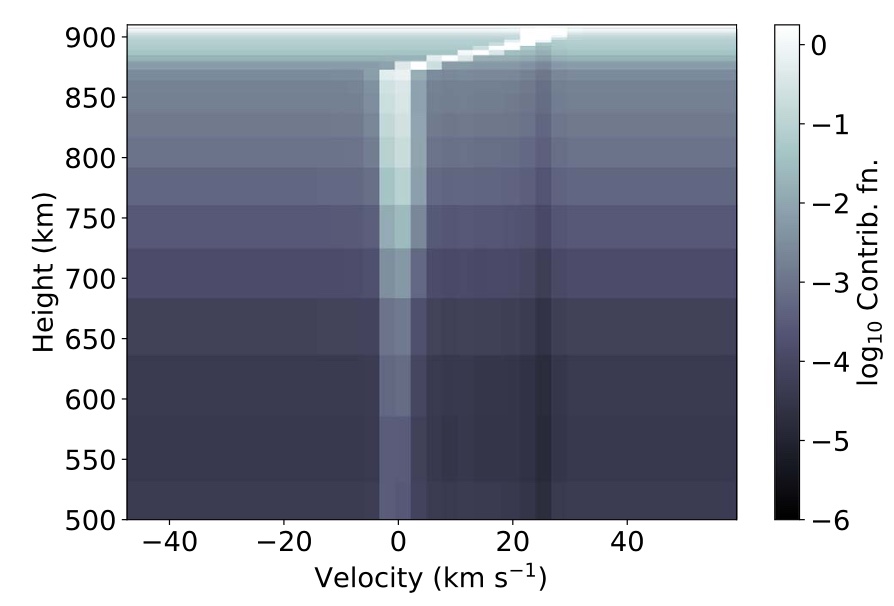}% This is a *.eps file
\end{center}
\caption{The contribution function to the emergent intensity of \ion{Fe}{ii} 2814.45~\AA. Integrating through height yields the emergent intensity. Note the intense, but narrow, condensation in the upper chromosphere producing bright, redshifted satellite component alongside the stationary component. Figure from \cite{2020ApJ...895....6G}. \copyright AAS. Reproduced with permission.}\label{fig:contrib}
\end{figure}

Modelling the \ion{Fe}{ii} lines, including averaging the synthetic spectra over the IRIS $\tau_{exp} = 2.4$~s exposure time, revealed that this condensation did indeed result in a satellite component  forming, that subsequently decelerated towards the stationary component as the condensation pushed deeper. Synthetic satellite components are shown in the righthand side of Figure~\ref{fig:satellite}, where colour represents time in the simulation. Figure~\ref{fig:contrib} shows the contribution function to the emergent intensity (i.e. where the line forms) of the \ion{Fe}{ii} line. The bright contribution from redshifted material is obvious, as is the narrowness of the condensation appearing in the upper chromosphere. From the formation properties of the lines, such as Figure~\ref{fig:contrib}, we can understand the origin of the satellite components. Flare heating in the chromosphere enhances the lines, but in a region without meaningful mass flows so that the near-stationary component is bright. Once the condensation develops at the top of the chromosphere/base of transition region and becomes dense it begins to produce Doppler shifted emission from those same ions. Since the stationary components are relatively narrow the very redshifted emission appears as a separate line. As the condensation accrues mass while it propagates deeper, it slows, such that the Doppler shift of the satellite component reduces and it merges with the stationary component, taking on the red-wing asymmetry appearance. The optical thickness of the lines comes into play as those lines that are very optically thick and form higher in altitude (e.g. the \ion{Mg}{ii} or \ion{C}{ii} resonance lines) will more quickly see the merging of the stationary and redshifted components (with the whole line appearing redshifted in many cases). In those cases the large opacity means that little light can escape the top of the condensation once enough density is accrued, and so the stationary component is not visible. For lines with smaller opacity both components can be seen. Further, the large opacity broadening of the resonance lines means that the redshifted component would likely not appear as a fully separate component, rather as a red wing asymmetry. 

The intensity of both the surrounding continuum and the stationary components agreed with the observations, as did the magnitude of the Doppler shifts. However, several discrepancies did exist. The satellite component was much too intense, outshining the stationary component, and was much too narrow (as discussed in Paper 2  \citealt{kerr2022_irisflarereview_p2} of this review line widths are a perennial problem). The onset and evolution of the satellite components also occurred on a more aggressive timescale compared to the observations, though not egregiously so (a factor $2-3\times$ faster). Finally, varying the low-energy cutoff or energy flux illustrated how sensitive the properties and evolution of the condensation are. For example, increasing the energy flux to the upper range of the estimates in this particular case produced a condensation much too fast compared to the observations with a larger-than-observed continuum intensity. Similarly, increasing the low-energy cutoff meant that the condensation developed too late and was too slow. This demonstrates the possibility that such observations can be used to guide or constrain the range of plausible electron beam parameters consistent with the evolution of UV radiation in a particular flare (though I stress `guide'; I do not believe we are yet in a position to use them independent of X-ray observations).

\cite{2020ApJ...895....6G} explicitly demonstrated how useful high-cadence spectral characteristics are in confirming the inferred properties of the electron beam. They also demonstrated why we should pay attention to weaker lines in addition to the more commonly studied resonance lines. Similar condensations were shown in the models discussing the much broader \ion{Mg}{ii} h \& k lines \cite[][]{2016ApJ...827..101K,2019ApJ...883...57K,2019ApJ...885..119K} but in those cases instead of producing satellite components, they produced small asymmetries in the red wings. In comparison to the resonance lines of \ion{Mg}{ii} and \ion{C}{ii}, \ion{Fe}{ii} 2814.45~\AA\ has a much lower opacity, probing more easily the deeper layers of the chromosphere \citep[][]{2017ApJ...836...12K}. Earlier modelling of condensation timescales found a similar discrepancy in the timescales compared to H$\alpha$ observations \citep{1989ApJ...346.1019F} as those noted by \cite{2020ApJ...895....6G}. IRIS's very high spatial resolution suggests that the answer to this discrepancy does not lie in the superposition of flows from many unresolved elements (though note that the difference in timing is only a factor 2 or so, not the order of magnitude that is the case for evaporative upflows!). An over-dense condensation could explain why the modelled satellite components were brighter than the stationary components. Though not focussing on red wing asymmetries, \cite{2019ApJ...883...57K} shows \ion{Mg}{ii} 2791~\AA\ lines with a redshifted satellite component that is weaker than the stationary component. In that simulation the condensation is not very dense, hence the smaller intensity. A means to obtain an estimate of the electron density from the broadening of high-order Balmer lines was presented by \cite{2022ApJ...928..190K}, and the effects of improved Stark broadening are now included in \radyn\ \& \rh. Coordinated DKIST \& IRIS observations of the Balmer lines and FUV/NUV spectra would shed light on this discrepancy and condensation densities \citep[see also the comprehensive discussion regarding model-data discrepancies in][Section 5]{2022ApJ...928..190K}.

\subsection{Flows in Small Scale Heating Events}
It is not yet known if the physics of flares scales from the very large (M and X class events) to the very small (micro or nanoflares), but it is a reasonable assumption that electrons could be accelerated even in small events \citep[see recent evidence from NuSTAR observations of (sub-) microflares][]{2020ApJ...891L..34G,2021MNRAS.507.3936C}. Rapid variations in Doppler motions (both blue- and redshifts) and intensities have been observed at the base of coronal loops in the transition region, leading \cite{2014Sci...346B.315T} and \cite{2018ApJ...856..178P} to investigate via \radyn\ modelling if non-thermal electron distributions injected into the transition region from the corona could explain these `nanoflare' signatures.  The total energy deposited was estimated as being $6\times10^{24}$~erg \citep[based on][]{1988ApJ...330..474P}, compared to $10^{30-32}$~erg for moderate-to-large flares. This equated to an energy flux of $1.2\times10^{9}$~erg~s$^{-1}$~cm$^{-2}$ considering the area of the footpoint emission, and an assumed dwell time of $10$~s based on lifetimes of short-lived brightenings in transition region moss. Note that the total energy, and the energy flux, associated with the nanoflares observed by \cite{2014Sci...346B.315T} and \cite{2018ApJ...856..178P}, while fairly small, is not so different from an individual flare loop; it is the much greater number of flare loops / greater flare volume that leads to the larger total energy.

\cite{2018ApJ...856..178P} performed a parameter study sampling different non-thermal electron distributions, and injected those electrons into two pre-flare atmospheres, one initially cool and tenuous (a low density corona at $1$~MK), and one hotter and dense (a high density corona at $3$~MK). The latter represents an active region loop, the former being more quiet Sun-like, and both had extended plateaus of higher temperatures in the chromosphere compared to VAL-C type atmospheres, maintained in the model by artificial non-radiative heating. The electron energy spectra had low energy cutoffs in the range $E_{c} = [5,10,15]$~keV. The low-energy cutoff, $E_{c}$, had a strong impact on the subsequent dynamics. Flares with a small $E_{c}$ efficiently heated the corona, driving the whole transition region to greater column mass, and therefore resulting in \ion{Si}{iv} redshifts, along with explosive evaporation. The corona is very tenuous, so it is easy to drive fast flows even at these low energy fluxes if $E_{c}$ is low such that energy is largely deposited in the upper transition region/lower corona. Larger values of $E_{c}$ means that there are relatively more high energy electrons that thermalise somewhat deeper, so that gentle evaporation through the transition region occurred and upflows at the temperatures that form \ion{Si}{iv} were present. Initially dense loops resulted in lower intensity emission and slower flows compared to initially tenuous loops. Loop length was also an important factor for the dense loops, with longer loops resulting in a larger portion of electrons thermalising in the corona.

\begin{figure}[h]
\begin{center}
\includegraphics[width=\textwidth]{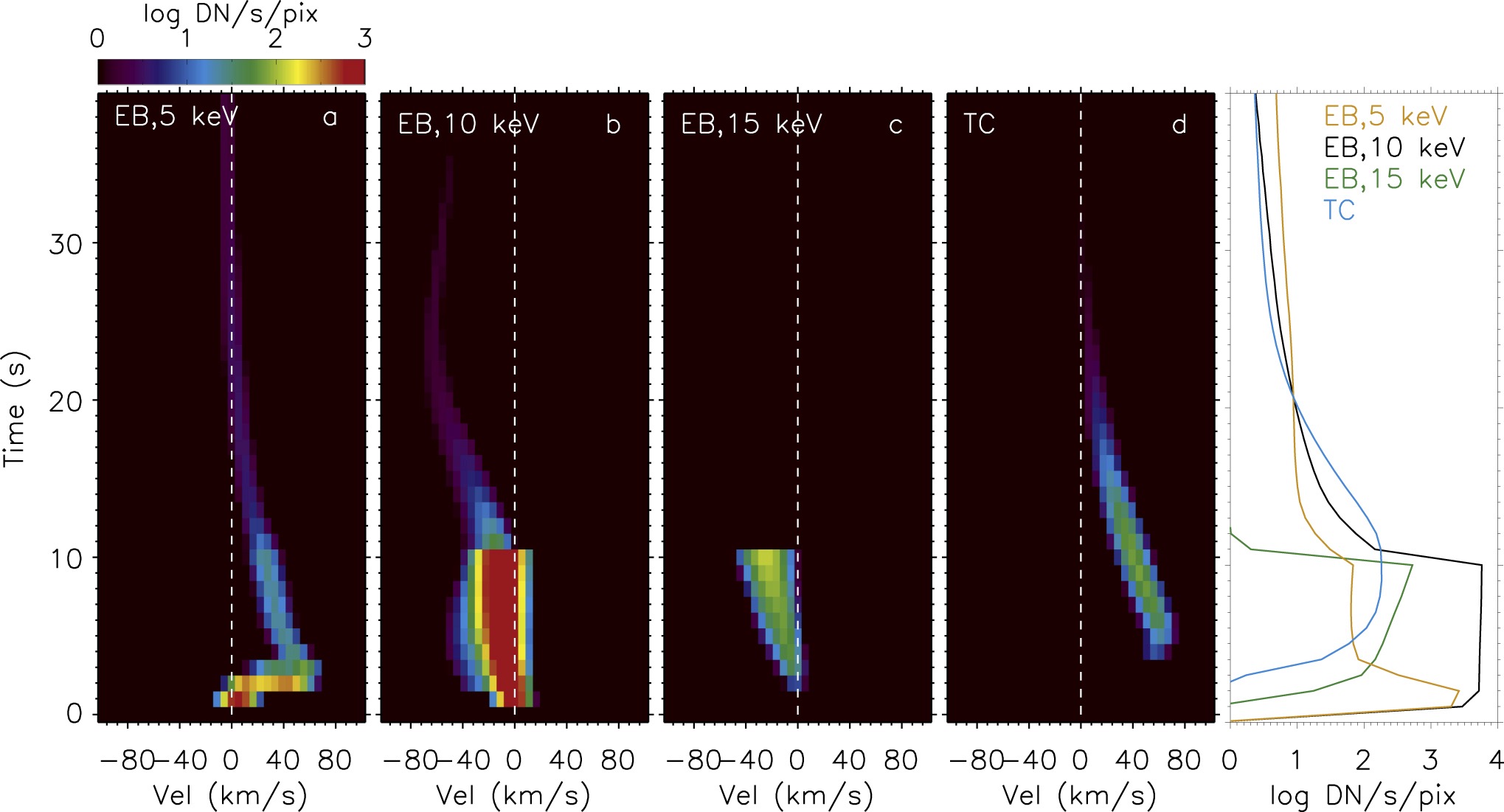}% This is a *.eps file
\end{center}
\caption{Demonstrating that non-thermal electron beams can produce blueshifted emission with $E_{c}>10$~keV in nanoflare simulations using \radyn. Panels \textbf{A-C} show \ion{Si}{iv} spectra for electron beams with different $E_{c}$, panel \textbf{D} shows the spectra in the case of a conduction driven flare, and the rightmost panel shows the lightcurves from each flare. Figure from \cite{2018ApJ...856..178P}. \copyright AAS. Reproduced with permission.} \label{fig:nanoflare}
\end{figure}

The dynamics of the loops had a direct impact on the synthetic emission, allowing a straightforward comparison to observations. \cite{2018ApJ...856..178P} synthesised \ion{Si}{iv} resonance line emission, degrading the spectra to IRIS resolution and count rates. Tenuous loops result in a very intense \ion{Si}{iv} response for all simulations, but reaching a peak in the $10$~keV experiment. For the dense loops, the softest non-thermal electron distribution ($E_{c} = 5$~keV) did not produce an appreciable response of \ion{Si}{iv} on rapid timescales. Blueshifts were seen in simulations with $E_{c} > 10$~keV, whereas redshifts were seen in simulations with $E_{c} = 5$~keV, as illustrated in Figure~\ref{fig:nanoflare} that shows the \ion{Si}{iv} spectra and lightcurves for simulations with different values of $E_{c}$. Once the loops have filled following evaporation and become denser, electrons thermalise more easily in the corona where they cause further heating.  Redshifts were also seen in experiments in which the same magnitude of energy was deposited directly in the corona and allowed to conduct through to the lower atmosphere.  No blueshifts were seen in that scenario.

 \cite{2018ApJ...856..178P} speculate that based on their loop modelling, the observations of impulsive brightenings in the transition region at the base of coronal loops are consistent with energy input to an initially cold, tenuous loop. Since observations show both blue and redshifted emission, this suggests that conduction driven heating of loop footpoints at the transition region alone is not consistent with the observations. That is, IRIS observations combined with \radyn\ modelling suggest that particles are accelerated even in small-scale brightenings, contributing to active region heating.

\subsection{Exploring Correlations Between Flare-Induced Upflows and Downflows}
Since the development of condensations and evaporations during flares is closely tied to the properties of energy injection, there could exist correlations between these flows inferred from spectral lines. \cite{2019ApJ...871....2S} explored the potential of such correlations using IRIS data, and also via \radyn\ modelling of electron beam driven flares. Seven flares were selected that conformed to a strict set of criteria, including fast scans ($< 90$~s), flare ribbons crossing the slit, both simultaneous RHESSI observations with prominent non-thermal components \textsl{and} RHESSI hard X-ray sources that were co-spatial with the IRIS slit (to ensure that the spectra studied were more likely to arise from non-thermal electron precipitation), and a source no farther than $750^{\prime\prime}$ from disk centre to attempt to minimise projection effects. The Doppler motions of the \ion{C}{ii} 1334.5~\AA\  and \ion{Fe}{xxi} 1354.1~\AA\ lines were determined, sampling the cooler layers more likely to experience a condensation and those more likely to experience upflows into the corona. A centre-of-gravity approach was used to obtain the Doppler shift of the optically thick\footnote{Hence fitting with a Gaussian function is not appropriate} \ion{C}{ii} 1334.5~\AA\  Doppler shift, whereas Gaussian fitting was performed for the optically thin \ion{Fe}{xxi} 1354.1~\AA\ line. For both lines, a mask of only the flaring sources was created, and the mean Doppler shifts in the area of the hard X-ray sources measured. Additionally, the maximum shifts were recorded. 

These same lines were modelled from a grid of electron beam driven flare simulations produced by the F-CHROMA consortium using \radyn\footnote{\url{https://star.pst.qub.ac.uk/wiki/public/solarmodels/start.html}}. This grid samples a large parameter space of non-thermal electron distributions, that were injected into a VAL3C-like \citep[][]{1981ApJS...45..635V} pre-flare atmosphere. Energy was injected for $t=20$~s, in a triangular profile peaking at $t=10$~s. While most of the F-CHROMA grid represents relatively weak-to-moderate heating events, it is a very useful resource for studying flare processes and I encourage the reader to access it for their own research.

 \cite{2019ApJ...871....2S} selected 20 flares from this grid that loosely were consistent with the observationally derived non-thermal electron distribution properties. For the observational analysis of hard X-rays, the area selected to determine the energy flux into the chromosphere was the $50$~\% contour of the hard X-ray source. As I discuss in other sections, this very likely is an overestimate of the areas, and thus an underestimate of the actual energy flux. At a cadence of $1$~s, synthetic \ion{C}{ii} and \ion{Fe}{xxi} spectra were obtained from the \radyn\ atmospheres, using \rhpar\ and data from CHIANTI, respectively. In the latter case, \ion{Fe}{xxi} emission was summed through the full extent of the half-loop, so no geometric effects of extended emission regions were considered. The same Doppler shift metrics derived from the observations were then obtained from the models.

A statistical analysis revealed that there was no meaningful correlation between observed mean \ion{C}{ii} redshifts and non-thermal energy flux, but there was in the models. While still not very statistically significant, the observed maximum redshifts did show a correlation with the injected non-thermal energy flux. Similarly, there were no meaningful correlations between observed \ion{Fe}{xxi} blueshifts and non-thermal energy flux, but there were in the models. Further discrepancies existed also. Unlike the observations, the models predicted that in some cases there should be modest \ion{C}{ii} blueshifts. Observed flows of \ion{Fe}{xxi} were on the order $100$~km~s$^{-1}$, consistent with the range of reported values in other studies, but some of the models exceeded $500$~km~s$^{-1}$. The differences between the models and data suggest that either the flares being compared were not apples-to-apples (in the sense that the energy flux in the observed flares may have been underestimated due to the choice of using hard X-ray areas, and so larger than those used in the models), or that physics is missing from the models. 

As an example of the latter, recent work by \cite{2022ApJ...931...60A} to include the suppression of thermal conduction in \radyn\ simulations due to turbulence and non-local effects revealed slower upflows, and a shift in the altitude at which the transition region forms. The latter results in the switch from upflow to downflow occurring at a different temperature for a given set of electron beam parameters \citep[in this case, those observed by][]{2009ApJ...699..968M}. That study focussed on EUV observations, but future work will include IRIS observables, which may address the discrepancies discovered by \cite{2019ApJ...871....2S}.

%%%%%%%%%%%%%%%%%%%%%%%%%%%%%%%%%%%%%%%%%%%%%
%% SUMMARY
%%%%%%%%%%%%%%%%%%%%%%%%%%%%%%%%%%%%%%%%%%%%%

\section{Summary}\label{sec:conc}
In this first part of a two-part review of IRIS observations and flare loop modelling I have introduced the four main modern flare numerical models that have been used alongside IRIS data: \radyn, \hydrad, \flarix, and \preft\ (though these latter two feature more in part two of this review). As well as this I have given an overview of how we synthesise IRIS observables from those models. With those models and the high spatial-, spectral-, and temporal resolution observations provided by IRIS we have learned much about flare-induced mass flows, that in the standard flare model assumed energy transport via directed beams of non-thermal electrons. 

Though there are details to work out, recent modelling work performed in an attempt to explain long-lived mass flows induced in flare footpoints, at the scale of IRIS resolution elements ($0.3-0.4^{\prime\prime}$), has convincingly demonstrated that there is a seeming demand for continued energy deposition into each footpoint for up to several minutes. Multi-threaded modelling results also suggests that this energy deposition is not coherent; that is, not every thread is energised at the same time, different threads (of which there may be many hundreds within each $0.167^{\prime\prime}$ pixel) are energised at different times, and for different durations. There are questions persisting here. We do not know if electrons are accelerated, and subsequently thermalised within the same small volume of an IRIS pixel for many minutes, nor even how long this would occur for on a single thread. Hard X-ray sources currently lack the high spatial resolution obtainable at UV wavelengths, and also have historically had low dynamic range so that only the strongest sources are observable. Still, hard X-ray sources are seen to move over time in a flare and not always linger for a long duration in a single location, and strong hard X-rays are not observed in the gradual phase.

IRIS observations of redshifted satellite components and red-wing asymmetries have been successfully modelled using field-aligned loop models in which non-thermal electron beams drive strong condensations. These condensations are narrow, and dense, producing redshifted emission that slows over time. Again, there are details to work out. For example the ratio of the redshifted to stationary component is not well-captured in models compared to observations. This could suggest an over-dense condensation in the model compared to the observation. The disparity between timescales of the redshifted components in the models compared to the observations is not as stark as those in the upflow scenario; condensation timescales in the models are more aggressive than in the observations, but do not reach the more than an order of magnitude differences that the modelled upflows suffer from.

Curiously, this means that we have different solutions to each problem: multi-threaded models with continued energy release is required to understand upflow behaviours, whereas single loop models with suitable non-thermal electron beam parameters that produce condensations can explain the chromospheric downflows. Transition region flows (e.g. \ion{Si}{iv}) are somewhat of an in-between case, with some observations requiring multi-threaded observations, but individual bursts in other observations can be captured with single loop modelling (see also Paper 2  \citealt{kerr2022_irisflarereview_p2}). It is important that we now work to reconcile this seeming contradiction  between the need for multi-threaded versus single loop modelling for different parts of the atmosphere. One resolution may be to have continued energy deposition into the transition region and upper chromosphere but at a magnitude too weak to drive strong condensations. This was hinted at by the results of multi-threaded modelling which suggested that the intensity of \ion{Si}{iv} emission should decrease towards the end of the energy deposition phase, whereas it remained rather flat for as long as energy was injected to the \hydrad\ models (see Figure~\ref{fig:multithreaded}). Experiments with alternate forms of gradual phase energy deposition should be performed since there is not compelling evidence for non-thermal electrons towards the end of each pixels lifetime. Perhaps direct heating in the coronal portion of the loop following reconnection with a subsequent conductive heat flux carrying energy into the lower atmosphere. Of course, area expansion of loops and suppression of conduction via turbulence and non-local effects \citep[e.g.][]{2018ApJ...865...67E,2018ApJ...856...27Z,2022ApJ...931...60A} mitigate the energetic requirements to sustain the temperature and density (and therefore line intensity) through the gradual phase, which might also indicate that after some time the energy flux in multi-threaded models should be decreased. Determining the appropriate parameters of suppression of conduction and the area expansion factors should also, therefore, be a priority.

Continued observations with IRIS will help here, but we must also look to future observations. IRIS demonstrated the benefits of detailed spectroscopy at high spatial and temporal resolution, and while it does sample different regions of the atmosphere it does have a fairly sparse temperature coverage. The upcoming Solar-C/EUVST instrument will have capabilities comparable to or slightly better than IRIS, but with a substantially denser temperature coverage, and a higher standard cadence\footnote{IRIS has recently started performing $\sim1$~s cadence flare observations, targeting the strongest lines.}. Observations will be available from photosphere to corona, with several hot flare lines ($5-15$~MK). As such, EUVST is very well placed to perform detailed studies of mass flows during flares that encompass simultaneously the chromosphere, transition region, and corona (with each layer sampled by many lines). Comprehensive analyses such as those performed by \cite{2009ApJ...699..968M} and \cite{2022ApJ...936...85S} over this wide temperature range, and with higher spatiotemporal resolution, should be a priority to better understand the evolution of mass flows in flares. EUVST will be complemented by observations from the Multi-slit Solar Explorer \citep[MUSE;][]{2020ApJ...888....3D,2022ApJ...926...53C} which has a more limited temperature coverage, focussing on the corona and flare plasma, with one line sampling the transition region, but which has 37 slits, allowing imaging spectroscopy of an entire active region sized field of view to be performed in $<12$~s. MUSE observations of the flaring corona, covering the full flaring structure, will hopefully shed light on continued energy release in the post-impulsive phase. From the ground, the now-operating Daniel K. Inouye Solar Telescope (DKIST) will also provide coverage from photosphere through corona, with an unprecedented spatial resolution $\sim0.1^{\prime\prime}$, which could reveal fine-structure in flare footpoints that guides future multi-threaded modelling.

I would also like to note that most of the flare modelling studies discussed here and in Paper 2 use standard pre-flare atmospheres (e.g. VALC, radiative equilibrium with different apex temperatures), but we know that the chromosphere is not homogenous. Real efforts should be made to (1) determine the large-scale impact on flare-induced flows of the choice of starting atmosphere, and to (2) perform bespoke modelling of flare footpoints where the pre-flare atmospheres are constrained by the observed chromosphere and corona. One means to guide the latter is the exciting advances in spectral inversions, including the IRIS2 resource \citep[][]{2019ApJ...875L..18S}, in which machine learning techniques were used with the STiC inversion code \citep[][]{2019A&A...623A..74D} and IRIS data to allow quick inversions of IRIS \ion{Mg}{ii} data to obtain the atmospheric stratification (with updates to include other lines being actively worked on, A. Sainz Diaz \textsl{private communication}, 2022).

Finally, mass flows are but one manifestation of solar flares. In Paper 2 of this review \citep[][]{kerr2022_irisflarereview_p2} I go on to discuss other plasma properties, energy transport mechanisms, and future directions of flare modelling.

%%%%%%%%%%%%%%%%%%%%%%%%%%%%%%%%%%%%%%%%%%%%%
%% ENDMATTER
%%%%%%%%%%%%%%%%%%%%%%%%%%%%%%%%%%%%%%%%%%%%%

\section*{Conflict of Interest Statement}
The author declares that the research was conducted in the absence of any commercial or financial relationships that could be construed as a potential conflict of interest.

\section*{Author Contributions}

GSK performed the literature review and wrote the manuscript. 

\section*{Funding}
GSK acknowledges funding via a NASA ROSES Early Career Investigator Award (Grant\# 80NSSC21K0460), and the Heliophysics Supporting Research program (Grant\# 80NSSC21K0010). 

\section*{Acknowledgments}
IRIS is a NASA small explorer mission developed and operated by LMSAL with mission operations executed at NASA Ames Research center and major contributions to downlink communications funded by the Norwegian Space Center (NSC, Norway) through an ESA PRODEX contract. This manuscript benefited from discussions held at a meeting of International Space Science Institute team: “Interrogating Field-Aligned Solar Flare Models: Comparing, Contrasting and Improving,” led by Dr. G. S. Kerr and Dr. V. Polito. I also thank the following colleagues for their help, and patience, with answering questions related to \radyn, \preft, \flarix, and \hydrad: Dr. Joel Allred, Dr. Mats Carlsson, Dr. Adam Kowalski, Dr. Vanessa Polito, Dr. Dana Longcope, Dr. William Ashfield, Dr. John Unverferth, Dr. Stephen Bradshaw, Dr. Jeffrey Reep, Dr. Jana {Ka{\v{s}}parov{\'a}, Dr. Petr Heinzel, and Dr. Michal Varady. Finally, I am grateful to the referees, who's careful and thorough comments improved this review.

\bibliographystyle{frontiersinSCNS_ENG_HUMS} 

\bibliography{Kerr_Frontiers_IRISFlareReview_LoopModels}

\end{document}